\documentclass[aps,twocolumn]{revtex4}

\usepackage{graphics}
\usepackage{graphicx}
\usepackage{bm}
\usepackage{latexsym}
\usepackage{color, soul}
\usepackage{mathrsfs}
\usepackage{braket}
\usepackage{enumerate}
\usepackage[usenames,dvipsnames]{xcolor}
\usepackage[colorlinks=true,citecolor=blue,linkcolor=blue,urlcolor=blue,backref=true]{hyperref}
\usepackage{float}
\usepackage[raggedright,tight]{subfigure}  
\usepackage{amsmath}
\usepackage{amsfonts}
\usepackage{braket}
\usepackage[normalem]{ulem}
\usepackage{cancel}
\usepackage{mathtools}
\usepackage{appendix}

\begin{document}

\title{Sampling methods to describe superradiance in large ensembles of quantum emitters}

\author{Daniel Eyles$^{1,2}$}
\author{Emmanuel Lassalle$^2$} 
\author{Adam Stokes$^3$}
\author{Ram\'on Paniagua-Domínguez$^{2,4}$}
\author{Ahsan Nazir$^1$}

\small
\affiliation{$^1$Department of Physics and Astronomy, University of Manchester, Oxford Road, Manchester M13 9PL, United Kingdom}
\affiliation{$^2$Institute of Materials Research and Engineering, A*STAR (Agency for Science, Technology and
Research), Republic of Singapore}
\affiliation{$^3$School of Mathematics, Statistics, and Physics, Newcastle University, United Kingdom}
\affiliation{$^4$Instituto de Estructura de la Materia (IEM), Consejo Superior de Investigaciones Científicas, Serrano 121, 28006 Madrid, Spain}

\begin{abstract}
Superradiance is a quantum phenomenon in which coherence between emitters results in enhanced and directional radiative emission. Many quantum optical phenomena can be characterized by the two-time quantum correlation function $g^{(2)}(t,\tau)$, which describes the photon statistics of emitted radiation. However, the critical task of determining $g^{(2)}(t,\tau)$ becomes intractable for large emitter ensembles due to the exponential scaling of the Hilbert space dimension with the number of emitters. Here, we analyse and benchmark two approximate numerical sampling methods applicable to emitter arrays embedded within electromagnetic environments, which generally provide upper and lower bounds for $g^{(2)}(t,0)$. We also introduce corrections to these methods (termed \emph{offset corrections}) that significantly improve the quality of the predictions. The optimal choice of method depends on the total number of emitters, such that taken together, the two approaches provide accurate descriptions across a broad range of important regimes. This work therefore provides new theoretical tools for studying the well-known yet complex phenomenon of superradiance in large ensembles of quantum emitters.

\end{abstract}

\maketitle

\section{Introduction} \label{theory}
Quantum emitters can interact via shared electromagnetic (EM) modes, which modify their collective spontaneous emission. This interaction gives rise, for example, to superradiance, whereby emission becomes enhanced and directional 
\cite{Dicke_Superradiance_1, Dicke_Superradiance_2, Dicke_Superradiance_3, Dicke_Superradiance_4, Dicke_Superradiance_5, osti_7365050, Subradiance_1, Subradiance_2}. In free space, superradiance is typically observed for emitter separations smaller than the wavelength $\lambda = 2\pi c/\omega_0$, where $\omega_0$ is the transition frequency. This phenomenon has been observed in a variety of experiments \cite{Superradiant_Experiment_4, MicroCavity_Experiment, Superradiant_Experiment_2_Cavity, Superradiant_Experiment_5_Cavity, Superradiant_Experiment_1} and holds promise for enabling a wide range of quantum technologies, including superradiant lasers \cite{Superraidant_Laser, Superraidant_Laser2}, photonic quantum networks \cite{Photonic_Network, Photonic_Network2}, quantum communications \cite{Superradiance_Comms, Quantum_Internet}, quantum sensing \cite{Subradiant_Use_1, Subradiant_Use_2}, quantum memory \cite{Subradiant_Memory, Subradiant_Memory2}, quantum computation \cite{Quantum_Computation}, and quantum cryptography \cite{Quantum_Crypto, Quantum_Crypto2}.

Collective emission is commonly characterized using the second-order correlation function $g^{(2)}(t,\tau)$, which measures how the joint probability of detecting two photons separated by a delay $\tau$ compares with that expected for statistically independent photons \cite{G2_Significance}. The \textit{photon bunching} or superradiant regime is defined by $g^{(2)}(t,0)>1$. Conversely, if the likelihood of detecting a second photon is reduced following the emission of a first, then $g^{(2)}(t,0) < 1$, in which case emitted photons are said to be \textit{antibunched} \cite{Single_Photon_Sources, Single_photon_Sources_2}, corresponding to the \textit{photon antibunching}, or subradiant, regime. While quantum optical correlation functions such as $g^{(2)}(t,\tau)$ are powerful tools for understanding the emission properties of quantum systems both theoretically \cite{Garcia_Universiality, Garcia_Chain, Early_time_Superradiance} and experimentally \cite{g2_Experiment_1, g2_Experiment_2, g2_Experiment_3}, their evaluation becomes increasingly challenging for large numbers of emitters $N$, since the dimension of the emitter Hilbert space scales exponentially as $2^{N}$. One must therefore resort to approximate approaches, such as mean-field theory \cite{Mean_Field_1, Mean_Field_2, Mean_Field_3, Mean_Field_4}.  

Here, we consider an alternative approximation strategy based on numerical sampling methods (SMs). In these approaches, a fixed (and relatively small) number of randomly selected emitters is used to compute the exact dynamics (including all mutual interactions), and this procedure is repeated multiple times to obtain an averaged result. In the present context, this corresponds to averaging $g^{(2)}(t,0)$
over many realisations. Inspired by an existing pairwise sampling method~\cite{Gardiner_Thermal, Densely_Packed_Dipoles}, we introduce a complementary $m$-wise sampling method that, when used in conjunction with the pairwise approach, will often provide upper and lower bounds for $g^{(2)}(t,0)$. 


Although the $m$-wise approach is motivated by the pairwise method, it is not a trivial extension. The pairwise method treats only two-emitter clusters exactly and neglects all single-emitter contributions to $g^{(2)}(t,0)$, which leads to a systematic overestimation of correlations. In contrast, the $m$-wise method retains these single-emitter terms and captures higher-order clusters of size $m>2$, resulting in a systematic underestimate. Despite having been employed in several contexts, the pairwise method has not previously been benchmarked against exact solutions for the types of free-space arrays considered here, leaving its accuracy and limitations largely unexplored. One of the central aims of this work is therefore to benchmark both approaches for the first time, identify the regimes in which each method is reliable, and demonstrate how, when taken together, they can be used to estimate upper and lower bounds on $g^{(2)}(t,0)$. In addition, we introduce \emph{offset corrections} that significantly improve the quantitative accuracy of both sampling methods while preserving their bounding character.

The paper is organised as follows. In Sec.~\ref{Theory Section}, we introduce the theoretical framework in which the correlation functions of interest are expressed in terms of the Green’s function describing the EM environment. In Sec.~\ref{Sampling Methods Section}, we present the SMs used to calculate second-order correlation functions, briefly reviewing the pairwise approach of Refs.~\cite{Gardiner_Thermal, Densely_Packed_Dipoles} and detailing the complementary $m$-wise method. In Sec.~\ref{Method Comparison Section}, we benchmark these methods for two representative systems:~a fully inverted array of identical emitters in free space,  arranged in an $N = 8\times 8$ square lattice, for which exact closed-form expressions for the second-order correlation function are available; and a one-dimensional (1D) chain of $N=8$ coherently driven emitters that have relaxed to a steady state, for which exact numerical solutions can be obtained despite the absence of closed-form expressions. 
Building on these benchmarks, we then apply the SMs to a system for which exact results are not available, namely the steady state of a \emph{coherently driven} $N = 8\times 8$ square lattice of identical emitters. Finally, our conclusions are summarised in Sec.~\ref{Summary section}.

\section{Theory}\label{Theory Section}

\subsection{Master Equation}

We consider $N$ identical dipolar quantum emitters, modeled as two-level systems with ground and excited states $\ket{g_\mu}$ and $\ket{e_\mu}$, located at positions $\bm{R}_\mu$, and described by dipole moment operators $\hat{\bm{d}}_\mu = \bm{d}_\mu( \hat{\sigma}^-_\mu + \hat{\sigma}^+_\mu)$, where $\sigma_\mu^+ = \ket{e_\mu}\bra{g_\mu}$ and $\sigma_\mu^- = \ket{g_\mu}\bra{e_\mu}$ are the raising and lowering operators of the $\mu$th emitter. The transition frequency $\ket{e_\mu}\to \ket{g_\mu}$ is denoted $\omega_0$. We allow weak monochromatic laser driving of the $N$-emitter system at frequency $\omega_L$ with  Rabi frequency $\Omega$. In a frame rotating at $\omega_L$ the weak-coupling (Born-Markov-secular) Lindblad master equation governing the $N$-emitter density matrix $\rho(t)$ is given by~\cite{Master_Eqn_1, Master_Eqn_2}
\begin{align}
        \Dot{\rho}(t) =& -i (\omega_0-\omega_L) \sum_{\mu}^N \big[\sigma_\mu^+\sigma_\mu^-, \rho(t)\big] - i \sum_{\mu \not=\nu}^N \Delta_{\mu\nu} \big[\sigma_\mu^+\sigma_\nu^-, \rho(t)\big] \nonumber \\
        & + \frac{i}{2} \sum_\mu^N \big[ \Omega e^{-i\bm{k}_L \cdot\bm{R_\mu}} \sigma_\mu^+ + \Omega e^{i\bm{k}_L \cdot\bm{R_\mu}} \sigma_\mu^-] +{\cal D}(\rho),
    \label{master equation0}
\end{align}
with dissipator 
\begin{align}
{\cal D}(\rho) = \sum_{\mu,\nu}^N \gamma_{\mu\nu} \Big( \sigma_\mu^- \rho(t) \sigma_\nu^+ - \frac{1}{2}\big\{\sigma_\nu^+ \sigma_\mu^-, \rho(t) \}\Big),
\end{align}
where $\{\cdot,\cdot\}$ denotes the anticommutator. The dissipative and coherent coupling coefficients $\gamma_{\mu\nu}$ and $\Delta_{\mu\nu}$ are defined as
\begin{subequations}
    \begin{gather}
        \gamma_{\mu\nu} = \frac{2\mu_0 \omega_0^2}{\hbar} \operatorname{Im}\big[ \bm{d}^*_\mu\cdot\bm{G}(\bm{R}_\mu, \bm{R}_\nu, \omega_0)\cdot \bm{d}_\nu\big], \label{gamma}\\
        \Delta_{\mu\nu} = -\frac{\mu_0 \omega_0^2}{\hbar} \operatorname{Re}\big[ \bm{d}^*_\mu\cdot\bm{G}(\bm{R}_\mu, \bm{R}_\nu, \omega_0)\cdot \bm{d}_\nu\big],  \label{lamb}
    \end{gather}
\end{subequations}
where the Green's tensor $\bm{G}(\bm{r}, \bm{s}, \omega)$ is found by solving Maxwell's equations with appropriate boundary conditions and uniquely determines the electric field at position $\bm{r}$ due to a point source at $\bm{s}$. In what follows, numerical solutions to Eq.~(\ref{master equation0}) are obtained using the Python package QuTiP \cite{Qutip}. 

\begin{figure}[t]
\begin{minipage}{\columnwidth}
\begin{center}
\hspace*{-1mm}\includegraphics[width=0.98\textwidth]{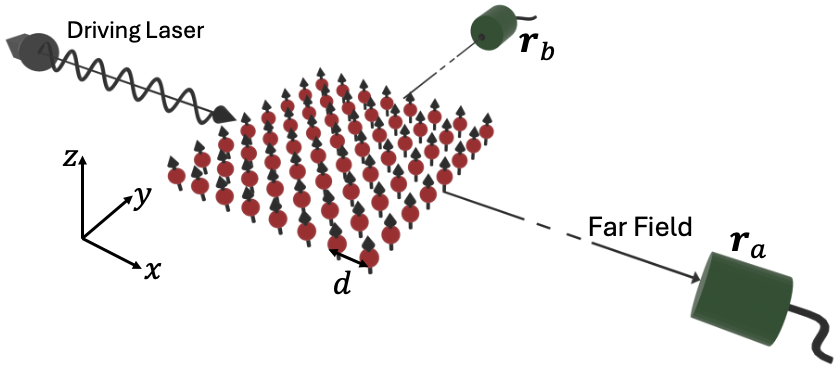}
\caption{
A square lattice of 64 emitters with lattice constant $d$ in the $x$--$y$ plane, coherently driven by a laser with wavevector $\mathbf{k}_L$ parallel to the $x$-axis, $\mathbf{k}_L = k_L \hat{\mathbf{x}}$ with $k_L=\omega_L/c$. All dipole moments are aligned parallel to the $z$-axis. Two detectors, labelled $a$ and $b$, are located in the far field of the lattice along the $x$- and $y$-directions, respectively.
}
\label{detector Diagram}
\end{center}
\end{minipage}
\end{figure} 

\subsection{General expressions for the two-time correlation functions}

The electric source-field of the $N$-emitter system can be written as ${\bm E}={\bm E}^{(+)}+{\bm E}^{(-)}$ where the positive frequency component is given by~\cite{Electric_Field}
\begin{equation}
    \hat{\bm{E}}^{(+)}(t,\bm{r}) = \mu_0 \omega_0^2 \sum_{\mu=1}^N \bm{G}(\bm{r}, \bm{R}_\mu, \omega_0) \cdot \bm{d}_\mu \sigma_\mu^-(t), \label{E complex greens function}
\end{equation}
and $\bm{E}^{(-)}(\bm{r}) = \bm{E}^{(+)}(\bm{r})^\dagger$. We consider a coincidence experiment involving two detectors, labelled $a$ and $b$, as depicted in Fig.~\ref{detector Diagram}. Detector $i=a,b$, located at position $\bm{r}_i$, has orientation labelled by $\alpha_i$, defined by the unit vector $\bm{e}_{\alpha_i}$. 

Following a first photon detection by detector $a$ at time $t$, the delay until a second detection event at detector $b$ is denoted by $\tau$, such that $\tau=0$ corresponds to coincident detections. The associated (elementary) second-order correlation function is defined by
\begin{equation}
        G^{(2)}(t,\tau) = \langle \hat{E}^\dagger_a(t) \hat{E}^\dagger_b(t+\tau) \hat{E}_b(t+\tau) \hat{E}_a(t) \rangle \label{G2}
\end{equation}
where $\hat{E}_i(t) = \bm{e}_{\alpha_i} \cdot \hat{\bm{E}}^{(+)}(t,\bm{r}_i)$. Rather than working directly with $G^{(2)}(t,\tau)$ it is common to introduce the normalised second-order correlation function
\begin{align}\label{g2pttau}
g_p^{(2)}(t,\tau) = {G^{(2)}(t,\tau) \over G^{(1)}_a(t)G^{(1)}_b(t+\tau)},
\end{align}
where $G^{(1)}_i(t) = \langle \hat{E}^\dagger_i(t) \hat{E}_i(t) \rangle$ is the first-order correlation function (intensity) associated with detector $i$.

Further useful normalised correlation functions can be defined from $g_p^{(2)}(t,\tau)$. In particular, by summing the first- and second-order correlation functions appearing in Eq.~(\ref{g2pttau}) over all detector orientations $\alpha_a$ and $\alpha_b$, one obtains an orientation-independent correlation function, denoted $g^{(2)}(t,\tau)$. By further integrating the detector positions $\bm{r}_a$ and $\bm{r}_a$ over enclosing surfaces $S_a$ and $S_b$, respectively, one arrives at a correlation function $\mathcal{G}^{(2)}(t,\tau)$ that is independent of both orientation and emission direction. As shown in Appendix \ref{Correlation function derivation}, all of these correlation functions can be written in the form
\begin{equation}
\begin{split}
    \mathcal{A}^{(2)}(t, \tau) &= \\
    & \frac{\sum_{\mu,\nu,\gamma,\epsilon}^N A^a_{\epsilon\mu} A^b_{\gamma\nu} \langle \hat{\sigma}_\mu^+(t) \hat{\sigma}_\nu^+(t+\tau) \hat{\sigma}_\gamma^-(t+\tau) \hat{\sigma}_\epsilon^-(t)\rangle }{\sum_{\mu,\nu,\gamma,\epsilon}^N A^a_{\nu\mu} A^b_{\epsilon\gamma}\langle \hat{\sigma}_\mu^+(t+\tau) \hat{\sigma}_\nu^-(t+\tau)\rangle \langle \hat{\sigma}_\gamma^+(t) \hat{\sigma}_\epsilon^-(t)\rangle} \label{A SOCF}
\end{split}
\end{equation}
where the coefficients $A^i_{\mu\nu}$ are defined by
\begin{equation}
    A^i_{\mu\nu} = 
    \begin{cases}
        G_{i\mu}G_{i\nu}^* & \quad \Rightarrow \quad \mathcal{A}^{(2)}(t,\tau) = g_p^{(2)}(t,\tau) \\
        \bm{G}_{i\mu}\cdot \bm{G}_{i\nu}^* & \quad \Rightarrow \quad \mathcal{A}^{(2)}(t,\tau) = g^{(2)}(t,\tau) \\
        \gamma_{\mu\nu} & \quad \Rightarrow\quad \mathcal{A}^{(2)}(t,\tau) = \mathcal{G}^{(2)}(t,\tau)
    \end{cases}
    \label{Coeff Matrix}
\end{equation}
with $G_{i\mu} = \bm{e}_{\alpha_i} \cdot \bm{G}\big(\bm{r}_i, \bm{R}_\mu, \omega)\cdot \bm{d}_\mu$, $\bm{G}_{i\mu} = \bm{G}(\bm{r}_i, \bm{R}_\mu, \omega) \cdot \bm{d}_\mu$, and $\gamma_{\mu\nu}$ given by Eq.~(\ref{gamma}). Eq.~(\ref{A SOCF}) 
constitutes a time-dependent generalisation, for arbitrary emitter number and system state, of related expressions previously derived in Refs \cite{Carminati_Paper, Wiegner_2015}. We emphasize that the expressions for $g_p^{(2)}(t,\tau)$ and $g^{(2)}(t,\tau)$ remain valid for arbitrary EM environments, while the expression for ${\cal G}^{(2)}(t,\tau)$ holds for any non-absorbing environment. 

For the purposes of introducing and analysing numerical methods applicable to large-$N$ systems, it suffices to consider free space, for which the Green's tensor takes the form \cite{Principle_Of_Nano_Optics}
\begin{equation}
\small
\begin{split}
    \bm{G}(\bm{r}, \bm{s}, \omega) = \bigg\{& \Big( \frac{3}{k^2 R^2} - \frac{3i}{kR} - 1\Big) \hat{\bm{R}}\hat{\bm{R}} \\
    &+ \Big(1 + \frac{i}{k R} - \frac{1}{k^2 R^2}\Big) \bm{I} \bigg\} \frac{e^{ikR}}{4\pi R},
\end{split}
\label{FS Greens Function}
\end{equation}
where ${\bm R} = \bm{r}-\bm{s}$, ${\hat{\bm R}}={\bm R}/R$, $\bm{I}$ is the unit dyad, and $k = \omega_0/c$. In this case, the single-emitter spontaneous emission rate is identical for all emitters, $\gamma_{\mu\mu} = \gamma_0$. A generalisation of the formalism presented here to higher-order correlation functions is given in Appendix~\ref{Correlation function derivation}.

\section{Application of sampling methods} \label{Sampling Methods Section}

\begin{figure}[t]
\begin{minipage}{\columnwidth}
\begin{center}
\hspace*{-1mm}\includegraphics[width=0.98\textwidth]{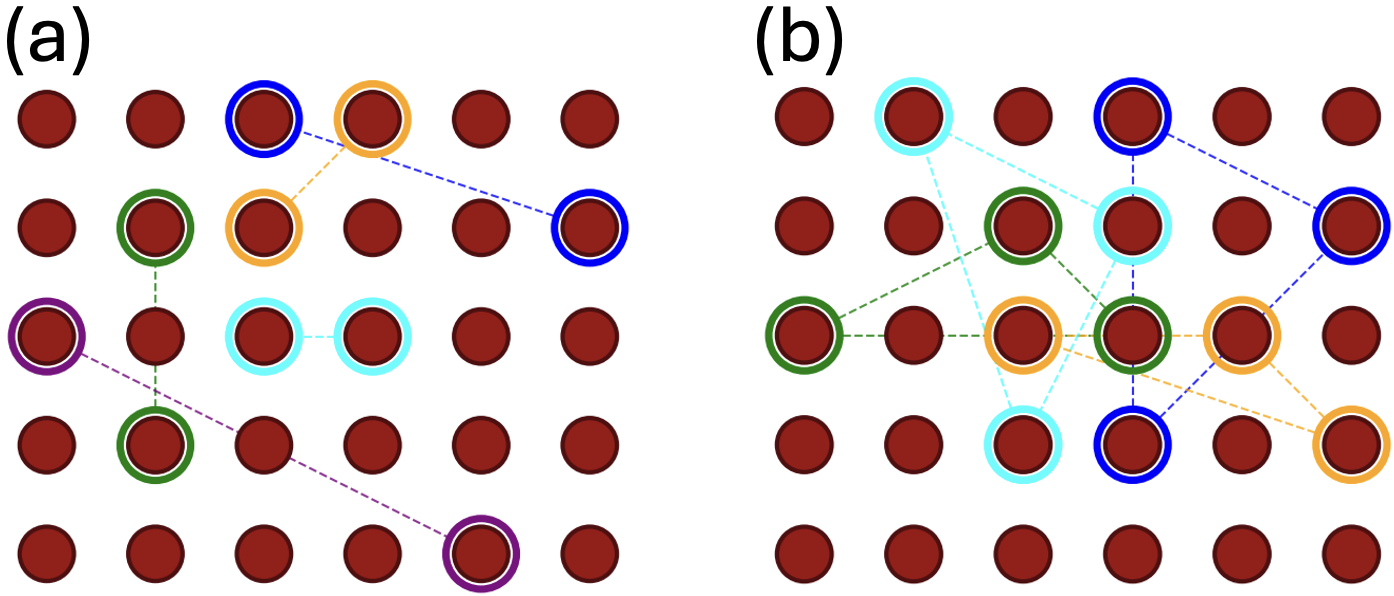}
\caption{
Schematic illustration of the approximate sampling methods.
(a) Pairwise sampling method with sample size $m = 2$ and total number of samples $S_2 = 5$.
(b) $m$-wise sampling method with sample size $m = 3$ and total number of samples $S_m = 4$.
Dashed lines indicate inter-emitter interactions characterised by the coefficients $\gamma_{\mu\nu}$ and $\Delta_{\mu\nu}$.
}
\label{Sampling Diagram}
\vspace*{-2mm}
\end{center}
\end{minipage}
\end{figure} 

Having established the theoretical framework, we now outline the pairwise and $m$-wise 
methods on which our analysis is based.

\subsection{Methods}
Recently, a numerical sampling technique suitable for modelling large ensembles of emitters coupled to EM modes has been proposed~\cite{Gardiner_Thermal}. In this approach, which we refer to as the \emph{pairwise sampling method}, random pairs of emitters are selected, the correlation functions for each pair are computed, and the results are then averaged over all samples. This procedure is illustrated schematically in Fig.~\ref{Sampling Diagram}(a). All single-emitter terms of the form $\langle \sigma_\mu^+ \sigma_\mu^+ \sigma_\mu^- \sigma_\mu^- \rangle$ and $\langle \sigma_\mu^+ \sigma_\mu^-\rangle\langle \sigma_\mu^+ \sigma_\mu^-\rangle$ that arise during the sampling procedure are neglected, in order to avoid overcounting single-emitter contributions when averaging. This approximation is justified provided that $N$ is sufficiently large such that the contribution of the 
single-emitter terms appearing in Eq.~(\ref{A SOCF}) is negligible. Under these conditions, it is proposed that accurate estimates of correlation functions can be obtained using the pairwise method, given a sufficiently large number of samples $S_2$.

Here, we introduce a complementary technique that significantly extends the range of regimes that can be treated accurately. 
Specifically, we consider random samples consisting of $m$ emitters, where $m$ may in principle take any value satisfying $1<m<N$, as depicted in Fig.~\ref{Sampling Diagram}(b). In contrast to the pairwise method, we retain single-emitter terms of the form $\langle \sigma_\mu^+\sigma_\mu^-\rangle\langle \sigma_\mu^+\sigma_\mu^-\rangle$, which allows for more accurate predictions in regimes where such contributions cannot be neglected. Owing to the unavoidable overcounting of these terms during the sampling process, one can expect to obtain a lower bound on the normalized correlation function sought. We refer to this technique as the \emph{m-wise sampling method}. Typically, the larger the total number of samples $S_m$, and the larger the size $m$ of samples, the greater the accuracy of the results found. In practice, the choice of $m$ must be balanced against the total number of samples that can be tractably computed given available computational resources. As will be shown in the following sections, the $m$-wise SM complements the pairwise approach, and together the two provide valuable insight into several important regimes of collective emission.

\subsection{m-wise convergence}
Within the $m$-wise method, there exists only a single parameter that can be tuned to alter predictions, namely the sample size $m$. 
It is therefore useful to explore the implications of varying $m$ within this method before undertaking a comparative analysis of both sampling approaches. To gain insight, it is optimal to work in the large $N$ regime; however, there are relatively few cases for which exact correlation functions can be computed. One such case is the so-called \textit{fully inverted array} in which all emitters are excited and undriven ($\Omega=0$). Without loss of generality, we assume that at time $t=0$ the density operator is $\rho(0) = \bigotimes_\mu^N \ket{e_\mu}\bra{e_\mu}$, and further restrict attention to coincident detection events by setting $\tau=0$. The required expectation values are then
\begin{subequations}
\begin{gather}
    \langle \sigma_\mu^+ \sigma_\nu^- \rangle = \delta_{\mu\nu}, \label{Inverted 2} \\
    \langle \sigma_\mu^+ \sigma_\nu^+ \sigma_\gamma^- \sigma_\epsilon^- \rangle = \delta_{\mu\gamma}\delta_{\nu\epsilon} + \delta_{\nu\gamma}\delta_{\mu\epsilon},
\end{gather}
\end{subequations}
which enables simplification of Eq.~(\ref{A SOCF}). In the present case, where $\mathcal{A}^{(2)}(0,0) = \mathcal{G}^{(2)}(0,0)$, this yields
\begin{equation}
    \mathcal{G}^{(2)}(0,0) = 1 + \frac{\sum_{\mu\not=\nu}^N \gamma_{\mu\nu}^2}{\left(\sum_\mu \gamma_{\mu\mu} \right)^2} - \frac{\sum_{\mu}^N \gamma_{\mu\mu}^2}{\left(\sum_\mu \gamma_{\mu\mu} \right)^2}.
    \label{eq:G20}
\end{equation}
This expression is tractable for all $N$, as the dependence on operators has been eliminated, and will therefore serve as a benchmark in what follows.

\begin{figure}[t]
\begin{minipage}{\columnwidth}
\begin{center}
\hspace*{-1mm}\includegraphics[width=0.95\textwidth]{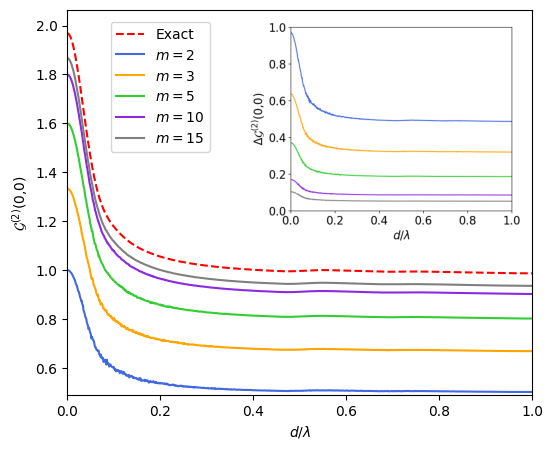}
\caption{$\mathcal{G}^{(2)}(0,0)$ as a function of normalized separation $d/\lambda$ defining a square lattice of $64$ identical emitters with all dipole moments aligned perpendicular to the plane of the lattice. The number of samples is $S_m=1000$. The exact result is shown by the red-dashed curve. All other curves are obtained using the $m$-wise SM for different sample sizes ($m$). The inset shows the difference between the exact method and those obtained using the $m$-wise SM, denoted $\Delta\mathcal{G}^{(2)}(0,0)$. }\label{Convergence analysis G2}
\end{center}
\end{minipage}
\end{figure}

We begin by considering a square lattice of 64 identical emitters in the $x$-$y$ plane, with fully aligned dipole moments perpendicular to the plane (along the $z$-axis) as shown in Fig.~\ref{detector Diagram}. We apply the $m$-wise SM to evaluate $\mathcal{G}^{(2)}(0,0)$, which is plotted as a function of the normalized emitter separation $d/\lambda$ in Fig.~\ref{Convergence analysis G2} (consistent results are obtained for the direction-dependent second-order correlation function $g^{(2)}(0,0)$, and are given in Appendix~\ref{g2 Appendix}~\footnote{We note also that very similar results (not presented) are obtained for the correlation function $g_p^{(2)}(0,0)$.}). As expected, the $m$-wise method produces an underestimate of the exact prediction that converges toward the latter as $m$ is increased, until coincidence occurs at $m=N$. This convergence is non-linear and becomes progressively slower as 
$m$ increases. The difference between the approximate and exact results (Fig.~\ref{Convergence analysis G2}, inset) decreases from its maximum at $d=0$ to its minimum as $d\to \infty$. This systematic behaviour can be exploited to construct a corrective offset, as discussed in the next section.
Fig.~\ref{Convergence analysis G2} also shows that, for fixed sample number $S_m$, the level of sampling noise decreases as $m$ is increased. This occurs because the distribution of possible values of $\mathcal{G}^{(2)}(0,0)$ for a given number of samples becomes narrower as $m$ increases, as discussed in Appendix~\ref{Dist Appendix}. 

\begin{figure}[t]
\begin{minipage}{\columnwidth}
\begin{center}
\hspace*{-1mm}\includegraphics[width=0.95\textwidth]{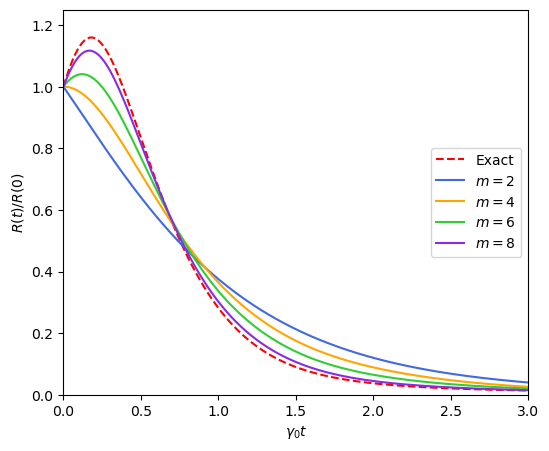}
\caption{Normalized emission rate as a function of time for a 1D chain of 9 identical emitters with dipole moments aligned perpendicular to the chain and separation $d=\lambda/10$. The exact result is shown in red, and all other curves are obtained using
the $m$-wise SM for different sample sizes ($m$) and $S_m = 1000$.}\label{Emission Rate}
\end{center}
\end{minipage}
\end{figure}

We next use the $m$-wise SM together with Eq.~(\ref{master equation0}) to predict the emission rate for $t>0$ of an undriven array that is fully inverted at $t=0$. Here, we consider a smaller system: a 1D chain of $N=9$ emitters with separation $d=\lambda/10$, where all dipole moments are aligned perpendicular to the chain. The emission rate is defined as the negative time derivative of the excited population $n_{\rm exc} = \sum_\mu^N \ket{e_\mu}\bra{e_\mu}$ expectation~\cite{Garcia_Chain, Early_time_Superradiance}:
\begin{equation}
    R(t) = - \frac{d\langle n_{\text{exc}}(t)\rangle}{dt} = \sum_{\mu\nu}^N \gamma_{\mu\nu} \langle \sigma_\mu^+(t) \sigma_\nu^-(t)\rangle,
\end{equation}
where $\gamma_{\mu\nu}$ is given by Eq.~(\ref{gamma}). As shown in Appendix~\ref{Inverted Array Derviation}, for the undriven system the initial gradient ${\dot R}(0)$ is related to the correlation function $\mathcal{G}^{(2)}(0,0)$ by~\cite{Early_time_Superradiance}
\begin{equation}
    \dot{R}(0) = R(0)^2\big(\mathcal{G}^{(2)}(0,0) - 1\big), \label{Emission Rate G2 relation} 
\end{equation}
which is positive in the superradiant regime. The emission rate as a function of time is plotted in Fig.~\ref{Emission Rate}. Its initial increase reflects enhanced collective emission due to constructive interference, while its subsequent decay indicates a transition to the subradiant regime caused by destructive interference and the formation of dark states \cite{Garcia_Universiality}.

Figs.~\ref{Convergence analysis G2} and~\ref{Emission Rate} show that the $m$-wise method with $m=2$ does not predict superradiance: the inverted state gives $\mathcal{G}^{(2)}(0,0)=1$ even at $d=0$. This is a limitation of $m=2$ sampling rather than a reflection of the underlying physics. Superradiant behaviour is only recovered when the $m$-wise method is applied with $m>2$. This restriction does not apply to the pairwise SM, since single-emitter contributions are neglected in that approach.

\subsection{Offset correction}

As discussed above, the difference between the predictions of the $m$-wise SM and exact results decreases monotonically with increasing emitter separation $d$, reaching a minimum in the limit $d\rightarrow\infty$. This behaviour suggests that it should be possible to improve the accuracy of the $m$-wise SM by applying a positive constant offset. A naive approach would be to first determine such offsets for a specific case, such as the inverted array, and apply them more generally. However, this would typically violate the useful property that the $m$-wise SM yields an underestimate of the true correlation function, thereby removing the guarantee that any predicted superradiance is genuine. Our aim is therefore to identify the maximum offset that preserves this lower-bound property. 

To this end, we consider the limit $d\to \infty$, in which inter-emitter interactions become negligible and the master equation admits the exact solution $\rho(t) = \rho_1(t)^{\otimes N}$, where $\rho_1(t)$ is the solution for $N=1$. In this independent-emitter regime, the decay matrix reduces to $\gamma_{\mu\nu} = \delta_{\mu\nu} \gamma_0$, and the normalized second-order correlation function is given by $\mathcal{G}_{\rm Ind}^{(2)}(0, 0) = (N-1)/N$, as derived in Appendix \ref{Inverted Array Derviation}. By contrast, in the same limit the prediction of the $m$-wise SM converges to $(m-1)/m$, such that the difference between the exact and approximate results is $1/m - 1/N$. Since this is the minimum difference for any $d$, it can be used as an offset correction that guarantees a lower bound is still obtained. A full derivation of this correction, along with further analysis, is given in Appendix~\ref{app:m-wise_key}. Predictions of the $m$-wise SM that include this correction will henceforth be labelled ``$m$-wise corr." For the pairwise method, the independent-emitter limit $d\to \infty$ yields $\mathcal{G}^{(2)}_{\text{Ind}}(0,0) = 1$. The offset correction defined in the same manner as for the $m$-wise SM is therefore $1-(N-1)/N = 1/N$, defining the pairwise corrected (``{pairwise corr.}") method (see Appendix~\ref{app:m-wise_key}).

\begin{figure}[t]
\begin{minipage}{\columnwidth}
\begin{center}
\hspace*{-1mm}\includegraphics[width=0.88\textwidth]{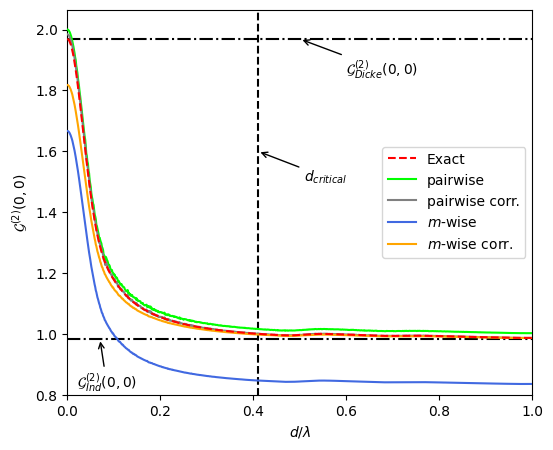}
\caption{$\mathcal{G}^{(2)}(0,0)$ as a function of the normalized emitter separation $d/\lambda$ for a fully inverted $8\times 8$ square lattice of identical emitters with dipole moments aligned perpendicular to the plane of the array. The exact result is shown by the red dashed curve. Predictions obtained using the pairwise and $m$-wise sampling methods are shown both with and without offset corrections, as indicated in the legend. Black dash-dotted lines specify the Dicke and independent emitter values. The critical separation $d_{\rm critical}$ for which $\mathcal{G}^{(2)}(0,0) = 1$ is specified by the vertical dashed line. For the pairwise method, we have chosen $S_2=10000$, and for the $m$-wise SM $m=6$ and $S_m=5000$.
}\label{Inverted Comparison G2}
\end{center}
\end{minipage}
\end{figure}

\begin{figure}[t]
\begin{minipage}{\columnwidth}
\begin{center}
\hspace*{-1mm}\includegraphics[width=0.95\textwidth]{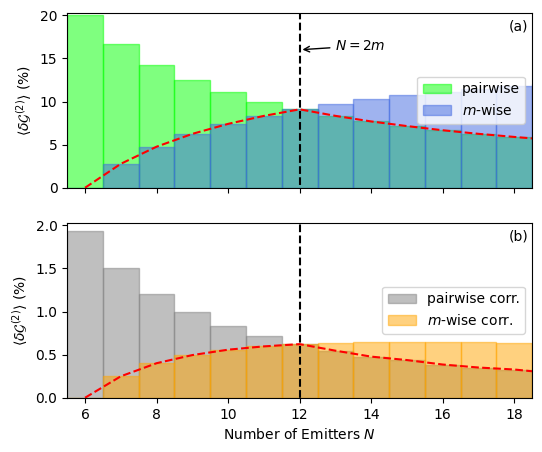}
\caption{
Mean percentage error $\langle \delta G^{(2)}_M \rangle$ as a function of the number of emitters $N$ for a 1D chain with dipole moments aligned perpendicular to the chain axis.
Results obtained (a) without offset corrections and (b) with offset corrections applied.
The red dashed line shows the minimum error across both methods and the vertical black line indicates the value $N = 2m$ where the optimal choice of method changes. For the pairwise SM we have chosen $S_2=10000$, and for the $m$-wise SM we take $m=6$ and $S_m=5000$.
}\label{Inverted Condition G2}
\end{center}
\end{minipage}
\end{figure}

\section{Sampling methods comparison} \label{Method Comparison Section}

We now turn our attention to comparing the $m$-wise and pairwise SMs across different systems and regimes. In all scenarios hereafter, we consider that at time $t=0$, the density operator is $\rho(0) = \bigotimes_\mu^N \ket{e_\mu}\bra{e_\mu}$.

\begin{figure*}[t]
    \centering
    \includegraphics[width=0.99\textwidth]{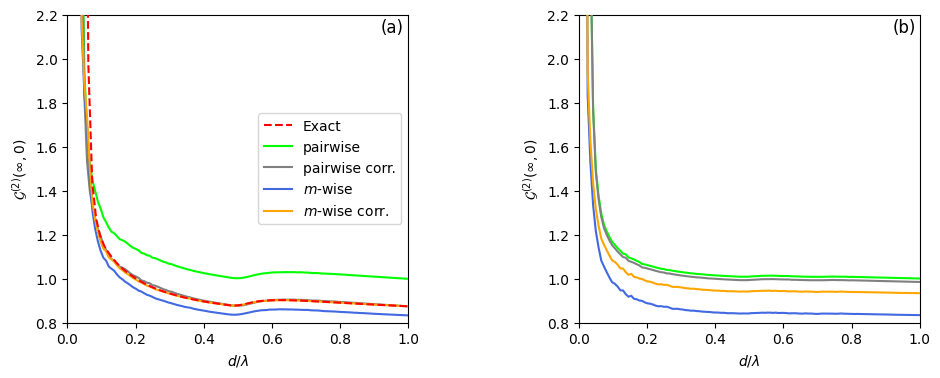}
\caption{
Steady-state second-order correlation function $\mathcal{G}^{(2)}(\infty,0)$ as a function of the normalized emitter separation $d/\lambda$ for coherently driven arrays.
(a) 1D chain of $N = 8$ emitters with dipole moments aligned perpendicular to the chain axis.
(b) Square lattice of $N = 64$ emitters with dipole moments aligned perpendicular to the plane of the array.
Predictions obtained using the pairwise and $m$-wise sampling methods are shown both with and without offset corrections, as indicated in the legend. The driving parameters were set at $\Omega=5\gamma_0$ and $\bm{k}_L = [1,0,0]$eV$^{-1}$, with $\omega_L=\omega_0 = 1$~eV. For the pairwise method, the number of samples taken was $S_2=10000$ and for the $m$-wise method, $m=6$ and $S_m=100$.
}\label{Steady Comparison G2}
\end{figure*}

\subsection{Inverted array comparison}

We begin by again considering the zero-delay second-order correlation function $\mathcal{G}^{(2)}(0,0)$ for a square lattice of 64 identical emitters that are all initially excited and undriven. The exact result, together with approximate predictions obtained using the pairwise and $m$-wise SMs, is shown in Fig.~\ref{Inverted Comparison G2}. From the figure, it is evident that the uncorrected pairwise SM significantly outperforms the uncorrected \(m\)-wise SM for the \(8 \times 8\) emitter array. Upon applying the corrections to both methods, their performances become comparable, with a marked improvement in overall accuracy in the $m$-wise case. At small separations, the \(m\)-wise SM, both uncorrected and corrected, continues to underpredict \(\mathcal{G}^{(2)}(0,0)\), which arises from an overestimation of the single-emitter contributions.

As shown in in Appendix \ref{Inverted Array Derviation}, at $d = 0$ one obtains the Dicke result $\mathcal{G}_{\rm Dicke}^{(2)}(0,0) = 2(N-1)/N = 2\times 63/64\simeq 1.97$. This maximal value depends on $N$ alone, and its upper bound is $\lim_{N\xrightarrow{}\infty}[\mathcal{G}_{\rm Dicke}^{(2)}(0,0)] = 2$. 
It is important to note that the Dicke result is specific to the inverted array case. In contrast, as $d\to \infty$ one finds that $\mathcal{G}_{\rm Dicke}^{(2)}(0,0) \to  \mathcal{G}_{\rm Ind}^{(2)}(0,0)=(N-1)/N$, which tends to unity when $N\to\infty$, while for $N=64$ it equates to $63/64\simeq 0.98$.

The exact prediction shows that the crossover between the superradiant regime $\mathcal{G}^{(2)}(0,0) >1$ (super-Poissonian photon statistics) and the subradiant regime $\mathcal{G}^{(2)}(0,0) < 1$ occurs when $d=0.41\lambda$, in agreement with the critical distance $d_{\rm critical}$ derived in Ref.~\cite{Garcia_Universiality}. The precise value of $d_{\rm critical}$ depends on the array geometry and on the orientations of the emitter dipole moments. It also increases with the number of emitters $N$ provided that the spatial dimensionality of the system is not altered with the addition of new emitters. Note that for 2D arrays with dipoles aligned perpendicular to the array as we have considered, $\mathcal{G}^{(2)}(0,0)$ does not always decrease monotonically with $d$, and revivals of superradiance can occur \cite{bettles2015cooperative, Garcia_Universiality}~\footnote{In the case considered here, a very small revival is seen at $d\approx 0.55\lambda$.}.

To assess the relative accuracies of the pairwise and $m$-wise SMs more systematically, we consider a 1D chain of emitters with dipole moments aligned perpendicular to the chain axis and allow the number of emitters $N$ to vary. Over the range $0\leq d/\lambda \leq 1$, we compute the mean percentage error
\begin{equation}
    \langle\delta \mathcal{G}^{(2)}_M \rangle = \frac{100}{n} \sum_{i=1}^n \frac{|\mathcal{G}^{(2)}(0,0;d_i) - \mathcal{G}_M^{(2)}(0,0;d_i)|}{\mathcal{G}^{(2)}(0,0;d_i)},
\end{equation}
where $n$ is the number of data points, $\mathcal{G}^{(2)}(0,0; d_i)$ denotes the exact value evaluated at separation $d_i$, and $\mathcal{G}_M^{(2)}(0,0; d_i)$ denotes the corresponding approximate value obtained using method $M=[2,\,m]$ for the pairwise and $m$-wise SMs respectively. 

Figs. \ref{Inverted Condition G2}(a) and \ref{Inverted Condition G2}(b) show these errors as a function of $N$, with and without the offset correction, respectively. We focus on a 1D geometry here, as it allows the progression of the error with increasing 
$N$ to be more clearly resolved; but an analogous behaviour is obtained for square lattices.  

As shown, the errors associated with the two methods exhibit opposite trends as 
$N$ increases: the error of the pairwise method decreases, while that of the 
$m$-wise method increases. This behaviour arises because the $m$-wise method overcounts single-emitter contributions, which are entirely neglected in the pairwise approach, leading to systematically lower predictions. The absolute accuracy of the $m$-wise method improves with increasing sample size $m$ and number of samples $S_m$ (consistent results are obtained for the direction-dependent second-order correlation function $g^{(2)}(0,0)$, and are given in Appendix~\ref{g2 Appendix}).

From this analysis of the 1D chain, we find empirically that the optimal method switches from the $m$-wise to the pairwise approach at the critical value 
$N=2m$. We have verified that the same transition point also applies to 2D arrays. To further sustain this, in Appendix~\ref{app:m-wise_key}, we derive this condition analytically in the particular case where emitters are located at the same position, for all correlation functions defined in Eq.~(\ref{A SOCF}).

Since the choice of $m$ must be balanced against the computationally feasible number of samples, the maximum tractable sample size $m_{\rm max}$ depends on available resources. For all systems considered here, arrays with $N < 2m_{\rm max}$ are more accurately described by the $m$-wise method, provided both $S_2$ and $S_m$ are sufficiently large to suppress sampling noise. For the data shown in Fig.~\ref{Inverted Condition G2}(a), the maximum error is approximately $10\%$, whereas in Fig.~\ref{Inverted Condition G2}(b) it is reduced to around $0.5\%$, demonstrating that the inclusion of offset corrections substantially improves the absolute accuracy. Applying the optimal sampling method for a given regime, together with these correction offsets, therefore yields accurate predictions of photon statistics.

\begin{figure*}[t]
    \centering
    \includegraphics[width=0.99\textwidth]{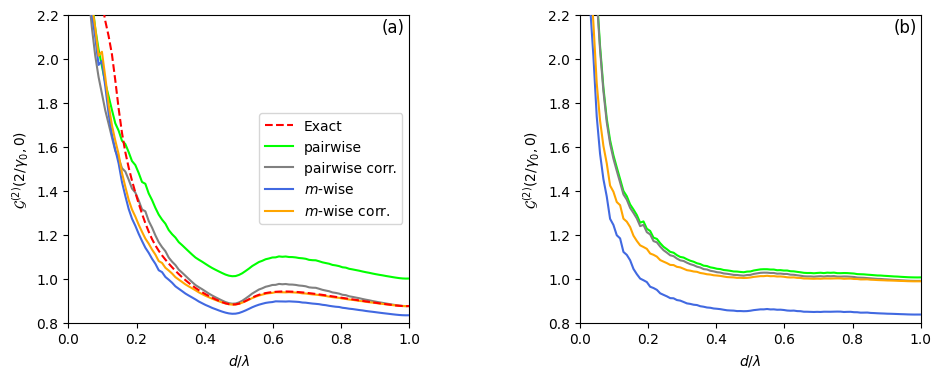}
\caption{
Second-order correlation function $G^{(2)}(2/\gamma_0,0)$ under free evolution ($\Omega = 0$) at time $t = 2/\gamma_0$ as a function of the normalized emitter separation $d/\lambda$.
(a) 1D chain of $N = 8$ emitters.
(b) Square lattice of $N = 64$ emitters. In both cases, the emitter dipole moments are aligned orthogonal to the array. 
Results are shown for both the pairwise and $m$-wise sampling methods, with and without offset corrections. For the pairwise method the number of samples taken was $S_2=10000$ and for the $m$-wise method, $m=6$ and $S_m=200$.
} \label{Finite Comparison G2}
\end{figure*}

\subsection{Coherently driven array}

Having established a framework for the pairwise  
and $m$-wise SMs, and benchmarked their performance in a numerically exact setting, we now consider more general states in which complex correlations between emitters arise.

We first focus on the steady state of a \emph{coherently driven} system, as depicted in
Fig.~\ref{detector Diagram}. Writing the master equation [Eq.~(\ref{master equation0})] in the form ${\dot \rho}={\cal L}(\rho)$, the steady state density matrix $\rho_{SS}$ is found by solving $\mathcal{L}\rho_{SS} = \bm{0}$ in Fock-Liouville space~\cite{Lindblad, F_L_Space}. Exact solutions can only be found for sufficiently small emitter numbers $N$. We therefore consider both a driven 1D chain of $N=8$ emitters extending along the $x$-axis, with dipole moments aligned parallel to the $z$-axis (which can be solved exactly), and a driven square lattice of $64$ emitters in the $x$-$y$ plane, again with all dipole moments oriented along the $z$-axis, for which an exact solution cannot be obtained. 

Fig.~\ref{Steady Comparison G2} shows predictions of the pairwise and $m$-wise SMs for both systems. For small emitter separations $d$
the steady-state correlation function $\mathcal{G}^{(2)}(\infty,0)$ diverges due to the coherent coupling term in the master equation, which contains the level shift $\Delta_{\mu\nu}$ (Eq.~(\ref{lamb})) that diverges at $d=0$. More generally, the observed behaviour is consistent with our earlier analysis of the critical emitter number $N=2m$, which marks the crossover between the regimes in which the two SMs are most accurate. 

For the 1D chain shown in Fig.~\ref{Steady Comparison G2}(a), where $N<2m$, the 
$m$-wise SM is qualitatively superior even in the absence of offset corrections. Including these corrections markedly improves the predictions of both methods. As in previous sections, the $m$-wise method consistently yields an underestimate of the true value, while the pairwise method typically produces an overestimate. We note, however, that this behaviour breaks down at small separations $d$. At these separations, collective behaviour is evidently not fully captured, even when offset corrections are included. This may result from the emergence of strong correlations at small separations, which, when combined with the neglect of single-emitter contributions, introduces inaccuracies.

The same trends are observed for the 64-emitter square lattice shown in Fig. \ref{Steady Comparison G2}(b). 
Although exact results are not available for comparison in this case, both SMs reproduce consistent qualitative features and, once offset corrections are included, are in good agreement over a wide range of separations. Deviations persist only at small $d$, where the discrepancy is likely associated with the divergence of parameters such as the dipole–dipole coupling, as discussed previously (for results obtained for the direction-dependent second-order correlation function $g^{(2)}(0,0)$, see Appendix~\ref{g2 Appendix}).

For completeness, we also consider the same systems undergoing free evolution ($\Omega = 0$). Predictions of $\mathcal{G}^{(2)}(2/\gamma_0, 0)$ obtained using both SMs are shown in Fig.~\ref{Finite Comparison G2}. As before, both approaches reproduce the correct qualitative behaviour for sufficiently large separations, with the $m$-wise method providing an underestimate and the pairwise method producing an overestimate, except for small separations $d/\lambda < 0.2$ (for results obtained for the direction-dependent second-order correlation function $g^{(2)}(0,0)$, see Appendix~\ref{g2 Appendix}).

\section{Summary} \label{Summary section}

We have developed and applied two numerical SMs to compute second-order correlation functions for large collections of dipolar emitters in arbitrary quantum states. In the pairwise method, proposed in~\cite{Gardiner_Thermal}, one takes random pairwise samples and ignores all single-emitter contributions. The second method introduced here, termed the $m$-wise method, employs random $ m$-sized subsampling of the emitter system and computes the total second-order correlation function for the subsample, retaining single-emitter terms. Contrary to the pairwise method, which typically produces an overestimate, the $m$-wise method systematically produces an underestimate. It is therefore generally more useful to apply both methods concurrently, thereby providing a bounded window of possible values of the correlation function, rather than considering either method alone. We find that for $m>N/2$ ($m<N/2$), the $m$-wise SM (respectively, the pairwise SM) is expected to be more accurate. For typical computational resources, the practical sample size for the $m$-wise SM is roughly of order $m\sim 10$.

Benchmarking predictions from both methods using the exactly-solvable case of a fully inverted array, the error in the optimal method, whether pair or $m$-wise, was found to be $\sim10$\%. We have shown, however, that this error can be reduced by an entire order of magnitude through the addition of constant offset corrections, $1/N$ and $1/m-1/N$ for the pairwise and $m$-wise methods, respectively. The quantity $1/m-1/N$ corresponds to the minimum difference between the exact prediction the $m$-wise result, and therefore maintains the useful property that this method should produce a lower bound. 

Taken together, these SMs provide both a consistency check on the qualitative features of photon correlation functions and a means of obtaining quantitatively accurate predictions. Their application should therefore prove valuable in regimes involving large emitter ensembles, where exact calculations become infeasible.

\bibliographystyle{apsrev4-2}
\bibliography{biblio}

\newpage
\newpage
\appendix

\onecolumngrid
\section{Derivation of the different correlation functions} \label{Correlation function derivation}
\subsection{Second-order correlation function}
In this section, we will outline the derivation of the different types of normalized correlation functions used in this paper. To begin, consider a system of $N$ emitters located at $\bm{R}_\mu$ for $\mu \in [1,N]$ and two detectors located at $\bm{r}_a$ and $\bm{r}_b$ with polarisations $\alpha_a$ and $\alpha_b$, which correspond to projections of the electric field along unit vectors $\bm{e}_a$ and $\bm{e}_b$, respectively.  Using this, we can define a normalized correction function as 
\begin{subequations}
    \begin{gather}
        g_p^{(2)}(t,\tau) = \frac{G^{(2)}(t,\tau)}{G_a^{(2)}(t)G_b^{(1)}(t+\tau)} \label{gp2}\\
        G^{(2)}(t,\tau) = \langle \hat{E}^\dagger_a(t) \hat{E}^\dagger_b(t+\tau) \hat{E}_b(t+\tau) \hat{E}_b(t) \rangle \label{G2} \\
        G^{(1)}_a(t) = \langle \hat{E}^\dagger_a(t) \hat{E}_a(t) \rangle. \label{G1}
    \end{gather}
\end{subequations}
which is dependent upon the emission direction and polarisation. From here, Eq.~\eqref{E complex greens function} for the electric field operator can be substituted to produce
\begin{equation}
\begin{split}
    g_p^{(2)}(t, \tau)|_{a,b} = \frac{\sum_{\mu,\nu,\gamma,\epsilon}^N G_{a\epsilon}G^*_{a\mu} G_{b\gamma}G^*_{b\nu}\langle \hat{\sigma}_\mu^+(t) \hat{\sigma}_\nu^+(t+\tau) \hat{\sigma}_\gamma^-(t+\tau) \hat{\sigma}_\epsilon^-(t)\rangle }{\sum_{\mu,\nu,\gamma,\epsilon}^N G_{a\nu}G^*_{a\mu} G_{b\epsilon}G^*_{b\gamma}\langle \hat{\sigma}_\mu^+(t+\tau) \hat{\sigma}_\nu^-(t+\tau)\rangle \langle \hat{\sigma}_\gamma^+(t) \hat{\sigma}_\epsilon^-(t)\rangle}. \label{gp}
\end{split}
\end{equation}
Where we have introduced the shorthand notation $G_{a\mu} = [\bm{e}_a \cdot \bm{G}(\bm{r}_a, \bm{R}_\mu,\omega_0)\cdot \bm{d}_\mu]$. This expression of the form of the equation defining $\mathcal{A}^{(2)}(t,\tau)$, and so we have obtained the first of the correlation functions. 
\\
From here it proves convenient to work in terms of $G^{(2)}(t,\tau)$ and $G_a^{(1)}(t)$. For the rest of these derivations, we will only show the work for $G^{(2)}(t,\tau)$ as the process applying to the denominator is identical. Now, we seek a correlation function that still considers directional emission but is not independent of polarization. To achieve this, we use the relation 
\begin{equation}
    \sum_{\alpha_a} G_{a\mu}G^*_{a\nu} = \bm{G}_{a\mu}\cdot\bm{G}^*_{a\nu}.
\end{equation}
Where, again we have introduce the short hand notation $\bm{G}_{a\mu} = \bm{G}(\bm{r}_a, \bm{R}_\mu,\omega_0)\cdot \bm{d}_\mu$. So, summing over all possible polarisation's $\alpha_a$ and $\alpha_b$ the numerator becomes 
\begin{equation}
    \sum_{\alpha_a} G^{(2)}(t,\tau) = (\mu_0\omega_0)^N\sum_{\mu,\nu,\gamma,\epsilon}^N \bm{G}_{a\epsilon}\cdot\bm{G}^*_{a\mu} \bm{G}_{b\nu}\cdot\bm{G}^*_{b\nu} \langle \hat{\sigma}_\mu^+(t) \hat{\sigma}_\nu^+(t+\tau) \hat{\sigma}_\gamma^-(t+\tau) \hat{\sigma}_\epsilon^-(t)\rangle .
\end{equation}
A very similar equation can be obtained when applying this procedure to the denominator. Taking the ratio of these equations obtains the normalized correlation function 
\begin{equation}
\begin{split}
    g^{(2)}(t, \tau)|_{a,b} = \frac{\sum_{\mu,\nu,\gamma,\epsilon}^N \big(\bm{G}_{a\epsilon}\cdot\bm{G}^*_{a\mu} \big)\big(\bm{G}_{b\gamma}\cdot\bm{G}^*_{b\nu} \big)\langle \hat{\sigma}_\mu^+(t) \hat{\sigma}_\nu^+(t+\tau) \hat{\sigma}_\gamma^-(t+\tau) \hat{\sigma}_\epsilon^-(t)\rangle }{\sum_{\mu,\nu,\gamma,\epsilon}^N \big(\bm{G}_{a\nu}\cdot\bm{G}^*_{a\mu} \big)\big(\bm{G}_{b\epsilon}\cdot\bm{G}^*_{b\gamma}\big)\langle \hat{\sigma}_\mu^+(t+\tau) \hat{\sigma}_\nu^-(t+\tau)\rangle \langle \hat{\sigma}_\gamma^+(t) \hat{\sigma}_\epsilon^-(t)\rangle}. \label{g}
\end{split}
\end{equation}
This is the second correlation function defined earlier. This correlation function is dependent upon the direction of emission only. 
\\
Finally, we will now derive the final correlation function defined previously. This correlation function will characterize the photon statistics independent of the direction of emission. To achieve this, we must integrate the detector locations over surfaces enclosing the system; $S_2$ and $S_m$. To do this, we require the relation
\begin{equation}
    \int_{S_2} d\bm{r}_a  \Big[\bm{G}(\bm{r}_a, \bm{R}_\mu,\omega_0)\cdot \bm{d}_\mu\Big]\cdot\Big[ \bm{G}(\bm{r}_a, \bm{R}_\nu,\omega_0)\cdot \bm{d}_\nu\Big]^* = \frac{c}{\omega_0} \bm{d}_\mu \cdot \operatorname{Im}\big[ \bm{G}(\bm{R}_\mu,\bm{R}_\nu,\omega_0) \big] \bm{d}_\nu^* \propto \gamma_{\mu\nu}. 
\end{equation}
Note, this result is proportional to the cross decay rate $\gamma_{\mu\nu}$. So, in addition to summing over the polarizations $\alpha_a$ and $\alpha_b$, we now also integrate over the surfaces $S_2$ and $S_m$. This sequence of actions results in
\begin{equation}
\begin{split}
    \sum_{\alpha_a,\alpha_b} \int_{S_2} \int_{S_m} d\bm{r}_a d\bm{r}_b G^{(2)}(t,\tau) \propto \sum_{\mu,\nu,\gamma,\epsilon}^N \gamma_{\epsilon\mu} \gamma_{\gamma\nu} \langle \hat{\sigma}_\mu^+(t) \hat{\sigma}_\nu^+(t+\tau) \hat{\sigma}_\gamma^-(t+\tau) \hat{\sigma}_\epsilon^-(t)\rangle
\end{split}
\end{equation}
Again, a very similar result can be obtained when considering the numerator (where both terms will have identical coefficients, and so will cancel when taking the ratio). Combining these, it can be shown that the final correlation function is given by
\begin{equation}
    \mathcal{G}^{(2)}(t, \tau)= \frac{\sum_{\mu,\nu,\gamma,\epsilon}^N \gamma_{\epsilon\mu}\gamma_{\gamma\nu} \langle \hat{\sigma}_\mu^+(t) \hat{\sigma}_\nu^+(t+\tau) \hat{\sigma}_\gamma^-(t+\tau) \hat{\sigma}_\epsilon^-(t)\rangle }{\sum_{\mu,\nu,\gamma,\epsilon}^N \gamma_{\nu\mu}\gamma_{\epsilon\gamma}\langle \hat{\sigma}_\mu^+(t+\tau) \hat{\sigma}_\nu^-(t+\tau)\rangle \langle \hat{\sigma}_\gamma^+(t) \hat{\sigma}_\epsilon^-(t)\rangle}.
    \label{GG2}
\end{equation}
which is the final correlation function that characterizes the photon statistics for direction-independent emission. 

\subsection{m-order correlation functions}
In the prior section, the derivation of the normalized second-order correlation function was shown. However, the method employed is valid for deriving the m-order correlation functions, and in this section, this derivation will be shown. To start, we consider $N$ identical emitters, located at positions $\bm{R}_\mu$ and dipole moments $\bm{d}_\mu$ where $\mu \in [1,N]$. Additionally, we consider $m$ detectors in the far field located at positions $\bm{r}_{a_i}$ and polarisation $\alpha_{a_i}$ which correspond to projections of the electric field along unit vectors of $\bm{e}_{a_i}$, where $i \in [1,m]$. Using this, we define the unnormalized correlation function as
\begin{equation}
    G^{(m)} = \langle \prod_i^m E^\dagger_{a_i}(t+\tau_i) \prod_i^m E_{a_i}(t+\tau_i) \rangle \label{GM}
\end{equation}
Where $\hat{E}_a = \bm{e}_a \cdot \hat{\bm{E}}^{(+)}(\bm{r}_a)$ and $\hat{\bm{E}}^{(+)}(\bm{r})$ is as defined in equation 1 in the main text. The equation is given in this form to ensure it remains in normal form, and note that for the time delay values $\tau_1 = 0$. As we already have the form of the first-order correlation function, we can focus on $G^{(m)}$, and we first substitute the explicit form of the electric field operator of the first product term in Eq.~\eqref{GM}
\begin{equation}
    \begin{split}
        \prod_i^m E_{a_i} &= (\mu_0 \omega^2)^m\bigg( \sum_{\mu_1}^N G_{a_1 \mu_1}\sigma_{\mu_1}^-\bigg)\bigg( \sum_{\mu_2}^N G_{a_2 \mu_2}\sigma_{\mu_2}^-(t+\tau_2)\bigg) ... \bigg( \sum_{\mu_m}^N G_{a_m \mu_m}\sigma_{\mu_m}^-(t+\tau_m)\bigg) \\
        &= \sum_{\mu_1}^N \sum_{\mu_2}^N ... \sum_{\mu_m}^N \bigg( \prod_i^m  G_{a_i \mu_i}\sigma_{\mu_i}^-(t+\tau_i) \bigg).
    \end{split}
\end{equation}
So, using this form, it is evident that the second product has a near-identical form, so the m-order correlation function can be expressed as 
\begin{equation}
    G^{(m)} =(\mu_0 \omega^2)^{2m} \sum_{\mu_1}^N ... \sum_{\mu_m}^N \sum_{\nu_1}^N...\sum_{\nu_m}^N \bigg[  \bigg(\prod_i^m G_{a_i \mu_i}\bigg)\bigg(\prod_i^m G^*_{a_i \nu_i}\bigg) \langle \bigg(\prod_i^m \sigma_{\nu_i}^+(t+\tau_i)\bigg) \bigg(\prod_i^m \sigma_{\mu_i}^-(t+\tau_i)\bigg) \rangle \bigg].
\end{equation}
While correct, this is not in a form similar to what was obtained for the second-order correlation function, but this can be achieved through
\begin{equation}
    \begin{split}
         \bigg(\prod_i^m G_{a_i \mu_i}\bigg)\bigg(\prod_i^m G^*_{a_i \nu_i}\bigg) &= \Big( G_{a_1 \mu_1} G_{a_2 \mu_2} ... G_{a_m \mu_m} \Big)\Big( G^*_{a_1 \nu_1} G^*_{a_2 \nu_2} ... G^*_{a_m \nu_m} \Big)\\
         &= \Big( G_{a_1 \mu_1}G^*_{a_1 \nu_1}\Big)\Big( G_{a_2 \mu_2}G^*_{a_2 \nu_2}\Big)...\Big( G_{a_m \mu_m}G^*_{a_m \nu_m}\Big) \\
         &= \prod_i^m G_{a_i \mu_i} G^*_{a_i \nu_i}
    \end{split}
\end{equation}
and so the mth-order correlation function becomes 
\begin{equation}
    G^{(m)} = (\mu_0 \omega^2)^{2m} \sum_{\mu_1}^N ... \sum_{\mu_m}^N \sum_{\nu_1}^N...\sum_{\nu_m}^N \bigg[ \bigg(\prod_i^m G_{a_i \mu_i} G^*_{a_i \nu_i}\bigg)   \langle \bigg(\prod_i^m \sigma_{\nu_i}^+(t+\tau_i) \bigg)\bigg(\prod_i^m \sigma_{\mu_i}^-(t+\tau_i)\bigg) \rangle \bigg], \label{Gm}
\end{equation}
which is in a suitable form. So using a similar definition to that seen in  Eq.~\eqref{gp2} we obtain an explicit form of $g^{(m)}_p(t,\tau)$ as
\begin{equation}
    g^{(m)}_p(t,\bm{\tau}) = \frac{\sum_{\mu_1}^N ... \sum_{\mu_m}^N \sum_{\nu_1}^N...\sum_{\nu_m}^N \bigg[ \bigg(\prod_i^m G_{a_i \mu_i} G^*_{a_i \nu_i}\bigg)   \langle \bigg(\prod_i^m \sigma_{\nu_i}^+(t+\tau_i)\bigg)\bigg(\prod_i^m \sigma_{\mu_i}^-(t+\tau_i)\bigg) \rangle \bigg]}{\prod_i^m \bigg( \sum_{\mu_i}^N  \sum_{\nu_i}^N G_{a_i \mu_i} G^*_{a_i \nu_i} \langle \sigma_{\mu_i}^+(t+\tau_i) \sigma_{\mu_i}^-(t+\tau_i) \rangle \bigg)}.
\end{equation}
This is the first of the three correlation functions stated. Following the work done in the previous section, we can perform a summation over all detector polarisations to obtain $g^{(m)}(t,\bm{\tau})$. This summation results in 
\begin{equation}
    \sum_{\alpha_{a_1}}\sum_{\alpha_{a_2}}..\sum_{\alpha_{a_m}} G^{(m)} = (\mu_0 \omega^2)^{2m} \sum_{\mu_1}^N ... \sum_{\mu_m}^N \sum_{\nu_1}^N...\sum_{\nu_m}^N \bigg[ \bigg(\prod_i^m \bm{G}_{a_i \mu_i} \cdot \bm{G}^*_{a_i \nu_i}\bigg)   \langle \bigg(\prod_i^m \sigma_{\nu_i}^+(t+\tau_i) \bigg)\bigg(\prod_i^m \sigma_{\mu_i}^-(t+\tau_i)\bigg) \rangle \bigg].
\end{equation}
So, the m-order correlation function corresponding to polarisation-independent, directional emission is given by
\begin{equation}
    g^{(m)}(t,\bm{\tau}) = \frac{\sum_{\mu_1}^N ... \sum_{\mu_m}^N \sum_{\nu_1}^N...\sum_{\nu_m}^N \bigg[ \bigg(\prod_i^m \bm{G}_{a_i \mu_i}\cdot \bm{G}^*_{a_i \nu_i}\bigg)   \langle \bigg(\prod_i^m \sigma_{\nu_i}^+(t+\tau_i)\bigg)\bigg(\prod_i^m \sigma_{\mu_i}^-(t+\tau_i)\bigg) \rangle \bigg]}{\prod_i^m \bigg( \sum_{\mu_i}^N  \sum_{\nu_i}^N \bm{G}_{a_i \mu_i}\cdot \bm{G}^*_{a_i \nu_i} \langle \sigma_{\mu_i}^+(t+\tau_i) \sigma_{\mu_i}^-(t+\tau_i) \rangle \bigg)}.
\end{equation}
Finally, we perform the $m$ integrations over the detector positions, using the relation
\begin{equation}
    \oint_{S_2} d\bm{r}_a \bm{G}_{a \mu}\cdot \bm{G}^*_{a \nu} \propto \gamma_{\mu\nu},
\end{equation}
where $\gamma_{\mu\nu}$ is the decay rate defined in equation 6a in the main text. Using this relation on both the numerator and denominator, we can obtain the expression
\begin{equation}
    \mathcal{G}^{(m)}(t, \bm{\tau}) = \frac{\sum_{\mu_1}^N ... \sum_{\mu_m}^N \sum_{\nu_1}^N...\sum_{\nu_m}^N \bigg[ \bigg(\prod_i^m \gamma_{\mu_i\nu_i}\bigg)   \langle \bigg(\prod_i^m \sigma_{\nu_i}^+(t+\tau_i)\bigg)\bigg(\prod_i^m \sigma_{\mu_i}^-(t+\tau_i)\bigg) \rangle \bigg]}{\prod_i^m \bigg( \sum_{\mu_i}^N  \sum_{\nu_i}^N \gamma_{\mu_i\nu_i}
    \langle \sigma_{\mu_i}^+(t+\tau_i) \sigma_{\mu_i}^-(t+\tau_i) \rangle \bigg)}.
\end{equation}
where we define the delay vector as, $\bm{\tau} = (\tau_1=0, \tau_2, ... \tau_m)$. This is a generalized form of that seen for the second-order correlation function in Eq.~\eqref{GG2}.

\section{Direction-dependent second-order correlation function} \label{g2 Appendix}
In this appendix, we conduct an identical analysis to that seen in the main text, but do not consider the directional dependent correlation function $g^{(2)}(t,\tau)|_{x,y}$, where the directions of emission we consider are along the x and y-axis. The conclusions which can be drawn from this work are identical to those seen in the main text regarding the viability of both SMs. The resurgence of a maximum of $g^{(2)}(t,\tau)|_{x,y}$ periodically is analysed in Appendix~\ref{Inverted Array Derviation}. 

\subsection{Convergence testing}
\begin{figure}[H]
\begin{minipage}{\columnwidth}
\begin{center}
\hspace*{-1mm}\includegraphics[width=0.4\textwidth]{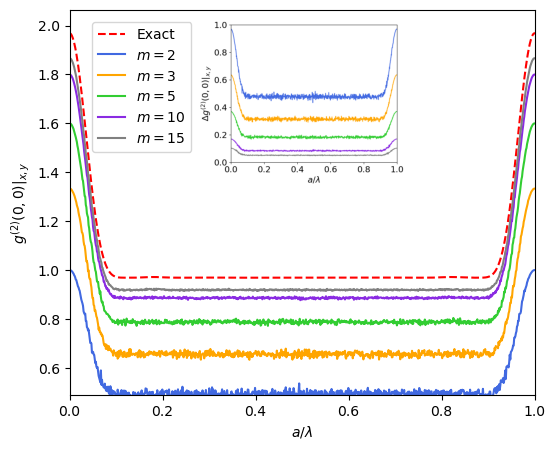}
\caption{Plots showing (a) how $g^{(2)}(0,0)|_{x,y}$ varies as a function of the emitter lattice constant $d$ for a system of 64 identical emitters arranged in a square lattice with all dipole moments aligned perpendicular to the plane of emitters. The number of samples taken was set to $S_m=1000$. The exact result is indicated in red, and the approximate solutions obtained using the $m$-wise SM are also shown for different sample sizes ($m$). }\label{Convergence analysis}
\end{center}
\end{minipage}
\end{figure}

\subsection{Inverted array comparison}
\begin{figure}[H]
\begin{minipage}{\columnwidth}
\begin{center}
\hspace*{-1mm}\includegraphics[width=0.75\textwidth]{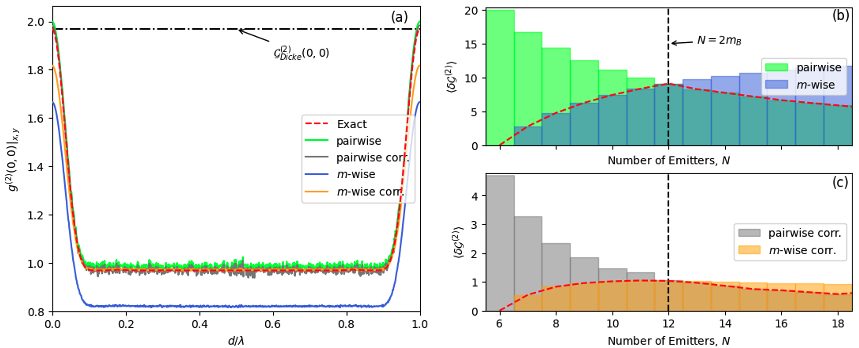}
\caption{Plots showing (a) the predictions of $g^{(2)}(0,0)|_{x,y}$ for both the pairwise and $m$-wise methods (Corrected and uncorrected) compared to the exact result for an inverted array of 64 emitters as explored in Fig. \ref{Convergence analysis}a. The Dicke result is indicated by the black horizontal dot-dashed line labeled $\mathcal{G}_{\rm Dicke}^{(2)}(0,0)|_{x,y}$. Plot (b) shows the mean percentage error in both methods as a function of the number of emitters, $N$, for a chain of emitters. The red dashed line here indicates the minimum error achieved by methods A and B (Uncorrected), and the vertical black line indicates the value of $N$ where the optimal choice of method changes, which is determined by the condition $N = 2m$. The number of random samples taken for the pairwise SM was set at $S_2=10000$, and for the $m$-wise SM the system size and number of random samples were set at $m=6$ and $S_m=5000$, respectively.}\label{Inverted Comparison}
\end{center}
\end{minipage}
\end{figure}

\subsection{Steady state comparison}
\begin{figure}[H]
\begin{minipage}{\columnwidth}
\begin{center}
\hspace*{-1mm}\includegraphics[width=0.75\textwidth]{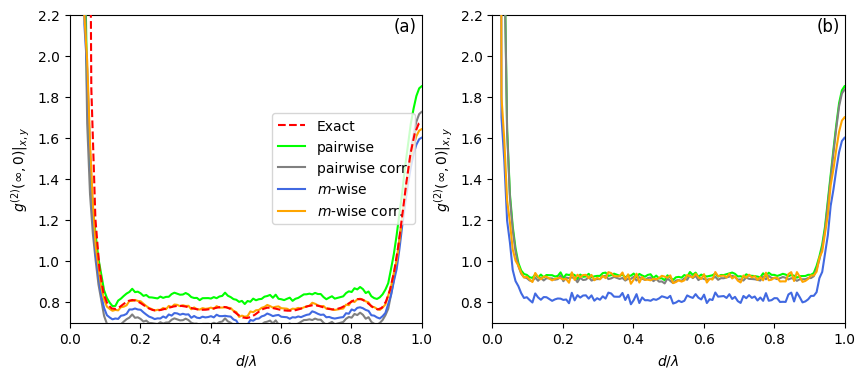}
\caption{Plots showing the predictions of $m$-wise and pairwise methods (corrected and uncorrected) for $g^{(2)}(\infty,0)|_{x,y}$ as a function of the emitter separation, $d$ for a system of coherently driven (a) eight emitters arranged into a 1D chain positioned along the $x$-axis and (b) 64 emitters arranged into a square lattice in the x-y plane. All emitter moments in both cases are aligned and oriented parallel to the $z$-axis and have been allowed to evolve to their steady state. The driving parameters were set at $\Omega=5\gamma_0$ and $\bm{k}_L = [1,0,0]$eV$^{-1}$, and the driving frequency is in resonance with the emitter transition frequency, $\omega_L=\omega_0 = 1$eV. For the pairwise SM, the number of samples taken was $S_2=10000,$ and for the $m$-wise SM, $m=6$ and $S_m=100$.  }\label{Steady Comparison g2}
\end{center}
\end{minipage}
\end{figure}

\subsection{Finite time comparison}
\begin{figure}[H]
\begin{minipage}{\columnwidth}
\begin{center}
\hspace*{-1mm}\includegraphics[width=0.75\textwidth]{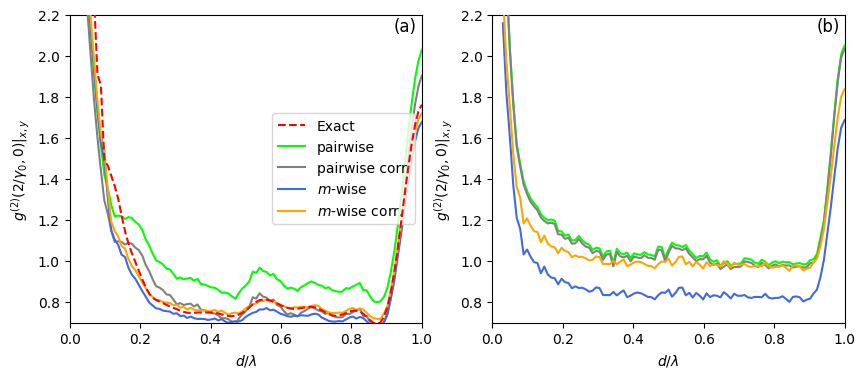}
\caption{Plots showing the predictions of the $m$-wise and pairwise methods (corrected and uncorrected) for $g^{(2)}(2/\gamma_0,0)|_{x,y}$ as a function of the emitter separation, $d$ for a system of (a) eight emitters arranged into a 1D chain positioned along the $x$-axis and (b)  64 emitters arranged into a square lattice in the x-y plane, that have freely evolved to the time $\gamma_0 t = 2$ under the master equation in Eq.~(\ref{master equation0}) with no driving ($\Omega=0$). All emitter moments in both cases are aligned and oriented parallel to the $z$-axis. For the pairwise SM, the number of samples taken was $S_2=10000$, and for the $m$-wise SM $m=6$ and $S_m=200$.  }\label{Finite Comparison}
\end{center}
\end{minipage}
\end{figure}

\section{Distribution of the sampling methods} \label{Dist Appendix}
\begin{figure}[H]
\begin{minipage}{\columnwidth}
\begin{center}
\hspace*{-1mm}\includegraphics[width=0.4\textwidth]{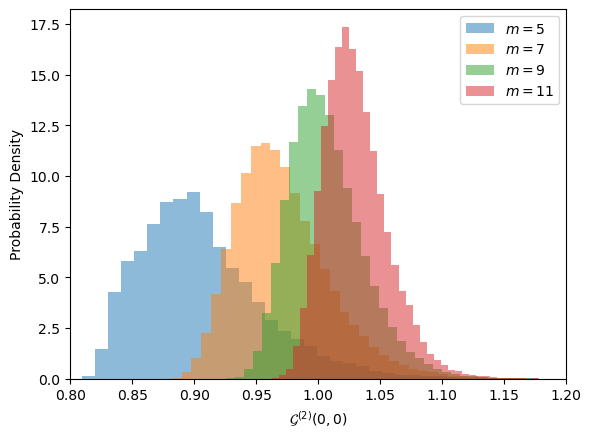}
\caption{Distributions of $\mathcal{G}^{(2)}(0,0)$ for the random sampling obtained using the $m$-wise SM for different sample sizes, $m$. The system considered is 121 emitters arranged into a 2D array with a spacing of $d = 0.1\lambda$. The resulting standard deviations for each [0.051, 0.038, 0.030, 0.025], respectively. }\label{Sampling Dist}
\end{center}
\end{minipage}
\end{figure}

As mentioned in the main text, when employing the $m$-wise SM, the required number of samples needed to reach convergence is reduced by increasing the sample size $m$. To see this, we can consider the standard error as the desired parameter to characterise the convergence of the mean value, and so, following this logic, we would expect the distributions of larger values of $m$ to possess a variance. To show this, the distributions of $\mathcal{G}^{(2)}(0,0)$ for different values of $m$ for the random samples of the $m$-wise SM can be seen in Fig. \ref{Sampling Dist}. This shows the distribution narrows as $m$ increases. 

\section{Inverted array photon statistics} \label{Inverted Array Derviation}
\subsection{Emission rate correlation function relation} \label{Emission Rate Correlation relation Appendix}
In this section, we will show the derivation of Eq.~\eqref{Emission Rate G2 relation} in the case of an undriven system ($\Omega=0$). To begin, we restate the definition of the emission rate for $N$ emitters
\begin{equation}
    R(t) = \sum_{\mu\nu}^N \gamma_{\nu\mu} \langle \sigma_\mu^+(t)\sigma_\nu^+(t)\rangle = \sum_{\mu\nu}^N \gamma_{\nu\mu} \text{Tr}\big[ \sigma_\mu^+\sigma_\nu^+\rho(t)\big],
\end{equation}
where we have transformed to the Schrodinger picture to obtain the third expression. Using this, we can determine the rate of change in the emission rate 
\begin{equation}
    \dot{R}(t) = \sum_{\mu\nu}^N \gamma_{\nu\mu} \text{Tr}\big[ \sigma_\mu^+\sigma_\nu^+\dot{\rho}(t)\big]. \label{R dot}
\end{equation}
Noting that the density matrix evolved according to the master equation seen in Eq.~\eqref{master equation0}, we can substitute the master equation into Eq.~\eqref{R dot}. To simplify this calculation we will split this into two components, $\dot{R}(t) = \dot{R}_{\text{c}}(t) + \dot{R}_{\text{i}}(t) $, where $\dot{R}_{\text{c}}(t)$, $\dot{R}_{\text{d}}(t)$ and $\dot{R}_{\text{i}}(t)$ correspond to the coherent, driving and incoherent terms, respectively. 

\subsubsection{Coherent terms}
Beginning with the coherent terms (which correspond to the first two terms in Eq.~\eqref{master equation0}), we obtain 
\begin{equation}
\begin{split}
    \dot{R}_{c}(t) = &-i(\omega_0-\omega_L) \sum_{ij}^N \gamma_{ij} \sum_{\mu}^N \Big( \text{Tr}\big[\sigma_i^+\sigma_j^- \sigma_\mu^+ \sigma_\mu^- \rho(t) \big] - \text{Tr}\big[\sigma_i^+\sigma_j^- \rho(t) \sigma_\mu^+ \sigma_\mu^- \big]\Big) \\
    &- i \sum_{ij}^N \gamma_{ij} \sum_{\mu\not=\nu}^N \Delta_{\mu\nu} \Big( \text{Tr}\big[\sigma_i^+\sigma_j^- \sigma_\mu^+ \sigma_\nu^- \rho(t) \big] - \text{Tr}\big[\sigma_i^+\sigma_j^- \rho(t) \sigma_\mu^+ \sigma_\nu^- \big]\Big) \\
\end{split}
\end{equation}
From these terms, it can be shown that through the use of the cyclical nature of the trace operator and swapping the index notation used on one of the terms present, these terms are identically 0, i.e. $\dot{R}(t) = 0$.

This result indicates that within an undriven system, the initial emission peak is independent of the coherent interactions and is only affected by driving, when there already exists coherence in the system.

\subsubsection{Incoherent terms}
We now consider the contribution from the dissipator to the rate of change of the emission rate. The contribution to the rate of change of the emission rate from this dissipator term is:

\begin{equation}
\dot{R}_{i}(t) = \sum_{\mu\nu}^N \gamma_{\mu\nu} \operatorname{Tr}\left[ \sigma_\mu^+ \sigma_\nu^- \mathcal{D}[\rho(t)] \right].
\end{equation}
Substituting the expression for the dissipator \(\mathcal{D}[\rho(t)]\), we get:

\begin{equation}
\dot{R}_{i}(t) = \sum_{ij}^N \gamma_{ij} \operatorname{Tr}\left[ \sigma_i^+ \sigma_j^- \left( \sum_{\mu, \nu}^N \gamma_{\mu\nu} \left( \sigma_{\mu}^- \rho(t) \sigma_{\nu}^+ - \frac{1}{2} \left\{ \sigma_{\nu}^+ \sigma_{\mu}^-, \rho(t) \right\} \right) \right) \right].
\end{equation}
Expanding this term results in the expression
\begin{equation}
    \dot{R}_{i}(t) = \sum_{ij}^N \sum_{\mu\nu}^N \gamma_{ij}\gamma_{\mu\nu} \Big( \operatorname{Tr}\big[\sigma_{i}^+ \sigma_{j}^- \sigma_{\nu}^- \sigma_{\mu}^+ \rho(t)\big] - \frac{1}{2}\big(\operatorname{Tr}\big[\sigma_{i}^+ \sigma_{j}^- \sigma_{\mu}^+ \sigma_{\nu}^- \rho(t)\big] + \operatorname{Tr}\big[\sigma_{i}^+ \sigma_{j}^- \rho(t)\sigma_{\mu}^+ \sigma_{\nu}^- \big] \big)  \Big).
\end{equation}
Again, in general, this term does not easily simplify, but under the assumption that the initial state is the fully inverted array, 
\begin{equation}
\begin{split}
    \dot{R}_{i}(0) &= \sum_{ij}^N \sum_{\mu\nu}^N \gamma_{ij}\gamma_{\mu\nu} \Big( \operatorname{Tr}\big[\sigma_{i}^+ \sigma_{j}^- \sigma_{\nu}^- \sigma_{\mu}^+ \rho(t)\big] - \frac{1}{2}\big( \delta_{\mu\nu} \delta_{ij} + \delta_{\mu\nu} \delta_{ij} \big)  \Big) \\
    &= \sum_{ij}^N \sum_{\mu\nu}^N \gamma_{ij}\gamma_{\mu\nu} \Big( \operatorname{Tr}\big[\sigma_{i}^+ \sigma_{j}^- \sigma_{\nu}^- \sigma_{\mu}^+ \rho(t)\big] \Big)  - R(0)^2\\
    & = R(0)^2 \mathcal{G}^{(2)}(0,0) - R(0)^2 = R(0)^2 (\mathcal{G}^{(2)}(0,0) - 1)
\end{split}
\end{equation}
Therefore, in the case of a fully inverted array, the initial rate of change in the emission rate $\dot{R}(0)$ is directly linked to the second-order correlation function via the relation 
\begin{equation}
    \dot{R}(0)^2 = R(0)^2 (\mathcal{G}^{(2)}(0,0) - 1).
\end{equation}
We emphasise that this derivation is valid only when there is no coherent driving applied to the system.

\subsection{Independent emitters}
To find the value at which $\mathcal{G}^{(2)}(0,0)$ converges to in the case of infinite separation is achieved by considering the set of emitters as independent of each other. In free space, this means the decoherent matrix can be expressed as $\gamma_{\mu\nu} = \gamma_0 \delta_{\mu\nu}$, where $\gamma_0$ is the spontaneous decay rate of an emitter and $\delta_{\mu\nu}$ is the Dirac delta. Using this when considering an inverted array, the correlation function can be expressed as 
\begin{equation}
\begin{split}
    \mathcal{G}_{\rm Ind}^{(2)}(0,0) &= \frac{\sum_{\mu,\nu,\gamma,\epsilon}^N \gamma_{\epsilon\mu}\gamma_{\gamma\nu}(1-\delta_{\mu\nu})(\delta_{\mu\gamma}\delta_{\nu\epsilon} + \delta_{\mu\epsilon}\delta_{\nu\gamma} )}{\sum_{\mu,\nu,\gamma,\epsilon}^N \gamma_{\nu\mu}\gamma_{\epsilon\gamma} \delta_{\mu\nu}\delta_{\gamma\epsilon}} \\
    &= \frac{\sum_{\mu,\nu,\gamma,\epsilon}^N \gamma_0^2 \delta_{\mu\epsilon}\delta_{\gamma\nu}(1-\delta_{\mu\nu})(\delta_{\mu\gamma}\delta_{\nu\epsilon} + \delta_{\mu\epsilon}\delta_{\nu\gamma} )}{\sum_{\mu,\nu,\gamma,\epsilon}^N \gamma_0^2 \delta_{\mu\nu}\delta_{\gamma\epsilon}} \\
    &= \frac{\sum_{\mu,\nu}^N (1-\delta_{\mu\nu})(\delta_{\mu\gamma}\delta_{\nu\epsilon} + \delta_{\mu\epsilon}\delta_{\nu\gamma} )}{\sum_{\mu,\gamma}^N 1 } \\
\end{split}
\end{equation}
By expanding the numerator, this expression simplifies to 
\begin{equation}
    \mathcal{G}^{(2)}(0,0) = \frac{N-1}{N}.
\end{equation}
This result clearly shows that when considering a set of independent emitters in free space, they will never show super-Poissonian photon statistics, and hence, reside in the subradiant regime.

\subsection{Dicke result}
Using the generalized notation introduced in equation 3 of the main text, we consider the system first explored by Dicke for $N$ identical emitters, all residing at the same position and in their excited state, $\rho = \ket{e}^{\otimes N}\bra{e}^{\otimes N}$. In this case, we can set $t=0$, without loss of generality, and $\tau=0$ to analyze the superradiant nature of this system. Under these considerations, the desired expectation values are given by 
\begin{subequations}
\begin{gather}
\langle \sigma_\mu^+ \sigma_\nu^- \rangle = \delta_{\mu\nu}, \label{Inverted 2} \\
\langle \sigma_\mu^+ \sigma_\nu^+ \sigma_\gamma^- \sigma_\epsilon^- \rangle = \delta_{\mu\gamma}\delta_{\nu\epsilon} + \delta_{\nu\gamma}\delta_{\mu\epsilon}.
\label{Inverted 4}
\end{gather}
\label{Inverted Relations}
\end{subequations}
where we have simplified notation by removing the functional dependence of the operators $\sigma_\mu^\pm (0) \equiv \sigma_\mu^\pm $. As we consider emitters at the same position and orientation, all elements of the coefficient matrix, defined in equation 4 of the main text, can be shown to be equivalent, $A_{\mu\nu} = A$. This equivalence is evident when stating the explicit form of the coefficients with all Green's functions evaluated at the same position. Therefore, substituting Eq.~\eqref{Inverted 2} and Eq.~\eqref{Inverted 4} into equation $\mathcal{A}^{(2)}(0,0)$ yields
\begin{equation}
\mathcal{A}_{Dicke}^{(2)}(0,0) = \frac{\sum_{\mu,\nu,\gamma,\epsilon}^N A^2 (\delta_{\mu\gamma}\delta_{\nu\epsilon} + \delta_{\nu\gamma}\delta_{\mu\epsilon}) }{\sum_{\mu,\nu,\gamma,\epsilon}^N A^2 \delta_{\mu\nu} \delta_{\delta\epsilon} }.
\end{equation}
The coefficients cancel out, resulting in sums over the Kronecker delta terms, giving us
\begin{equation}
\mathcal{A}^{(2)}_{Dicke}(0,0) = \frac{2(N-1)}{N}. \label{Dicke Result}
\end{equation}
This shows that the Dicke result is not specific to any of the correlation functions, and so applies to any normalized second-order correlation function evaluated for the case of the fully inverted array.

\subsection{$g^{(2)}(0,0)|_{a,b}$ position condition}
Within this section, we aim to derive conditions for emitter placements to observe maximal values in $g^{(2)}(0,0)|_{a,b}$ when considering an inverted array. We begin by obtaining the far-field Green's function. To begin, we take the far-field limit of the Free Space Green's function
\begin{equation}
    \lim_{r \xrightarrow{} \infty} \big[ \bm{G}(\bm{r}, \bm{s},\omega) \big] \approx \Big[ \bm{I} - \hat{\bm{r}}\hat{\bm{r}} \Big] \frac{e^{ikr}}{4\pi R} e^{-ik \hat{\bm{r}}\cdot\bm{s}}.
\end{equation}
where $\bm{r}$ denotes the detector location and $\bm{s}$ denotes emitter locations. For the systems considered in this work, we can assert all dipole moments are aligned ($\bm{d}_\mu = \bm{d}_\nu = \bm{d}$) in addition to imposing the condition that the emission of both photons is in the same direction ($\hat{\bm{r}}_a = \hat{\bm{r}}_b$). Using this, the coefficients (denoted as $A$ in the prior equations) become 
\begin{equation}
    A_{\mu\nu} = \bm{G}^*_{a\mu}\cdot \bm{G}_{a\nu} = \Big[\frac{e^{ikr_a}}{4\pi r_a} \bm{d} \cdot \Big( \bm{I} - \hat{\bm{r}}_a\hat{\bm{r}}_a \Big)\cdot \bm{d}\Big]^2 e^{ik \hat{\bm{r}}_a\cdot\bm{R}_\mu} e^{-ik \hat{\bm{r}}_a\cdot\bm{R}_\nu}
\end{equation}
From here, it is evident that the squared term is independent of the emitter position, and so for a given emission direction is constant. So, we will introduce the shorthand notation 
\begin{equation}
    \zeta_a = \Big[\frac{e^{ikr_a}}{4\pi r_a} \bm{d} \cdot \Big( \bm{I} - \hat{\bm{r}}_a\hat{\bm{r}}_a \Big)\cdot \bm{d}\Big]^2.
\end{equation}
Using this, the correlation function can be expressed as
\begin{equation}
   g^{(2)}(0,0)|_{a,b} = \frac{\sum^N_{\mu,\nu,\gamma,\epsilon} \zeta_a \zeta_b \langle\sigma_\mu^+\sigma_\nu^+\sigma_\gamma^-\sigma_\epsilon^-\rangle e^{ik \hat{\bm{r}}_a\cdot\bm{R}_\mu} e^{ik \hat{\bm{r}}_b\cdot\bm{R}_\nu} e^{-ik \hat{\bm{r}}_b\cdot\bm{R}_\gamma} e^{-ik \hat{\bm{r}}_a\cdot\bm{R}_\epsilon}}{\Big( \sum^N_{\mu,\nu} \zeta_a \langle \sigma_\mu^+ \sigma_\nu^- \rangle e^{ik \hat{\bm{r}}_a\cdot\bm{R}_\mu}e^{-ik \hat{\bm{r}}_a\cdot\bm{R}_\nu}\Big)\Big(\sum^N_{\mu,\nu} \zeta_b \langle \sigma_\mu^+ \sigma_\nu^- \rangle e^{ik \hat{\bm{r}}_b\cdot\bm{R}_\mu}e^{-ik \hat{\bm{r}}_b\cdot\bm{R}_\nu}\Big)},
\end{equation}
From here, it is evident that the coefficients $\zeta_{a/b}$ can be neglected and the correlation function relations for the inverted array can be substituted in. Additionally, we introduce the shorthand notation $\bm{R}_{\mu\nu} = \bm{R}_\mu - \bm{R}_\nu$. This expression now simplifies to 
\begin{equation}
   g^{(2)}(0,0)|_{a,b} = \frac{\sum^N_{\mu,\nu,\gamma,\epsilon}  (1-\delta_{\mu\nu})(\delta_{\mu\gamma}\delta_{\epsilon\nu} + \delta_{\mu\epsilon}\delta_{\gamma\nu})  e^{ik \hat{\bm{r}}_a\cdot\bm{R}_{\mu\epsilon}} e^{ik \hat{\bm{r}}_b\cdot\bm{R}_{\nu\gamma}} }{\Big( \sum^N_{\mu,\nu} \delta_{\mu\nu} e^{ik \hat{\bm{r}}_a\cdot\bm{R}_{\mu\nu}}\Big)\Big(\sum^N_{\mu,\nu} \delta_{\mu\nu} e^{ik \hat{\bm{r}}_b\cdot\bm{R}_{\mu\nu}}\Big)},
\end{equation}
Now the Kronecker delta terms can once again be expanded. The effects of the Kronecker deltas can be summarised into three aspects
\begin{itemize}
    \item $(1-\delta_{\mu\nu})$ Does not lead to cancellations.
    \item $\delta_{\mu\gamma}\delta_{\nu\epsilon}$ forces $\mu=\gamma$ and $\nu=\epsilon$.
    \item $\delta_{\mu\epsilon}\delta_{\nu\gamma}$ forces $\mu=\epsilon$ and $\nu=\gamma$ 
\end{itemize}
Following these rules, the correlation function can be significantly simplified to 
\begin{equation}
    g^{(2)}(0,0)|_{a,b} = \frac{1}{N^2}\sum_{\mu\nu}^N e^{ik \hat{\bm{r}}_a\cdot\bm{R}_{\mu\nu}} e^{ik \hat{\bm{r}}_b\cdot\bm{R}_{\nu\mu}} + 1 - \frac{2}{N}
\end{equation}
Due to the symmetric summation of the indices, this value is guaranteed to be real, but to maximise this value, the phase terms must equal 1 for all emitter pairs. To achieve this, the argument of these terms must be an integer multiple of $2\pi$. Additionally, we note that the wavenumber can be expressed in terms of the transition wavelength $k = 2\pi/\lambda$ and using these points, we obtain the condition 
\begin{equation}
    (\hat{\bm{r}}_a - \hat{\bm{r}}_b)\cdot (\bm{R}_{\mu}-\bm{R}_{\nu}) = n\lambda 
\end{equation}
where $n$ is an integer value. In this form, it is evident that the correlation function is always maximized in the case of $\bm{r}_a=\bm{r}_b$. However, in the case of emission in different directions, it is more convenient to express this condition in two separate equations
\begin{equation}
    \hat{\bm{r}}_a \cdot (\bm{R}_{\mu}-\bm{R}_{\nu}) = m\lambda \ \ \ \ \ \ \ \ \ \hat{\bm{r}_b} \cdot (\bm{R}_{\nu}-\bm{R}_{\mu}) = l\lambda,
\end{equation}
where $m$ and $l$ are again integer values. These equations make it clear that to maximize the value of $g^{(2)}(0,0)|_{x,y}$, the separation of the emitters along the direction of emission must be an integer multiple of the transition wavelength for the inverted array case.

\section{\textit{m}-wise SM key results}
\label{app:m-wise_key}
\subsection{\textit{m}-wise SM offset correction} \label{Offset Appendix}
\begin{figure}[H]
\begin{minipage}{\columnwidth}
\begin{center}
\hspace*{-1mm}\includegraphics[width=0.99\textwidth]{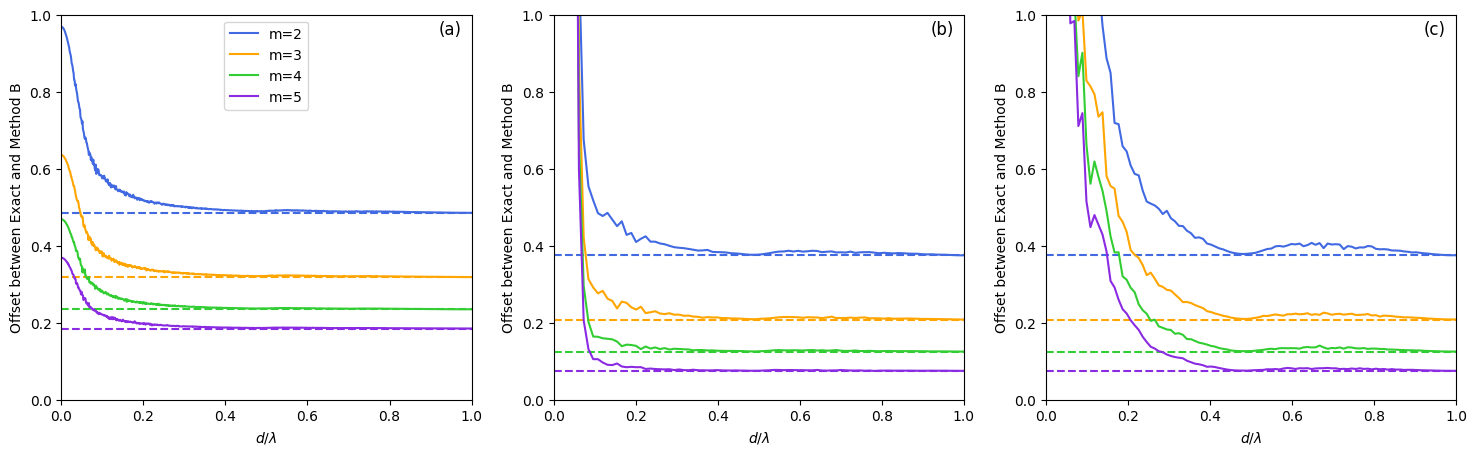}
\caption{Plots showing the offsets between the $m$-wise SM and the exact result for different sample sizes $m$ as a function of the emitter separation, $d$, for (a) The inverted array system seen in Fig. \ref{Inverted Comparison G2}a, (b) The steady state chain of emitters seen in Fig. \ref{Steady Comparison G2}a, and (c) the freely evolving system to time $\gamma_0t = 2$ seen in Fig. \ref{Finite Comparison G2}a. The dashed line corresponds to the difference in the independent emitter values $\mathcal{G}^{(2)}_{Ind}(t,0)|_N - \mathcal{G}^{(2)}_{Ind}(t,0)|_m = (N-1)/N - (m-1)/m$. }\label{Offset Plots}
\end{center}
\end{minipage}
\end{figure}

In this appendix, we will derive an analytical expression for the constant offset used to improve the accuracy of the $m$-wise SM. To achieve this, we must find the maximum correction value that ensures that the predictions of the $m$-wise SM remain an underestimate of the true value. To investigate this, Fig. \ref{Offset Plots} shows the offsets as a function of the lattice constant for the small emitter systems seen in Figs. \ref{Inverted Comparison G2}a, \ref{Steady Comparison G2}a, and \ref{Finite Comparison G2}a. 

From these plots, it becomes apparent that the maximum value which does not affect the underestimate behaviour corresponds to the independent emitter results. However, to implement this, we now require an approach to compute this offset for any system. To do this, we first note that for an independent system, the decay rates and coupling terms present in the master equation obey $\gamma_{\mu\nu} = \delta_{\mu\nu}\gamma_0$ and $\Delta_{\mu\nu} = 0$, respectively. As a result of this, the master equation becomes fully factorizable, meaning the total dynamics of the system can be computed through the single-emitter master equation. Under this, we let $\rho_1(t)$ be the arbitrary single emitter density matrix governed by the master equation 
\begin{equation}
    \begin{split}
        \dot{\rho}_1(t) =& -i (\omega_0-\omega_L) \big[\sigma^+\sigma^-, \rho_1(t)\big]  + \frac{i\Omega}{2}  \big[ e^{-i\bm{k}_L \cdot\bm{R}} \sigma^+ + e^{i\bm{k}_L \cdot\bm{R}} \sigma^-]  + \gamma_0 \Big( \sigma^- \rho_1(t) \sigma^+ - \frac{1}{2}\big\{\sigma^+ \sigma^-, \rho(t) \}\Big),
    \end{split}
    \label{master equation 1}
\end{equation}
which is analytically solvable. The total system density matrix is then defined as $\rho(t) = \rho_1(t)^{\otimes N}$. Knowing this significantly simplifies the calculation of $\mathcal{G}^{(2)}(t,0)$. To show this, let us consider an arbitrary density matrix 
\begin{equation}
    \rho_1(t) = \begin{pmatrix}
        p & c^* \\
        c & 1-p        
    \end{pmatrix},
\end{equation}
which can be considered as the desired solution to the single emitter master equation. We can now calculation the first-order correlation function $G^{(1)}(t)$ defined in Eq.~\eqref{G1} which gives
\begin{equation}
    \begin{split}
        G_{Ind}^{(1)}(t) = \sum_{\mu\nu}^N \gamma_{\mu\nu} \langle \sigma_\mu^+ \sigma_\nu^- \rangle &= \gamma_0 \sum_{\mu\nu}^N \delta_{\mu\nu} \operatorname{Tr}\big[ \sigma_\mu^+ \sigma_\nu^- \rho(t) \big] \\
        &= \gamma_0 \sum_{\mu}^N  \operatorname{Tr}\big[ \sigma_\mu^+ \sigma_\mu^- \rho(t) \big] = \gamma_0 \sum_{\mu}^N  \operatorname{Tr}\big[ \sigma_\mu^+ \sigma^- \rho_1(t) \big] \\
        &= \gamma_0 N p.
    \end{split}
\end{equation}
Following a similar process for $G_{Ind}^{(2)}(t,0)$, we obtain
\begin{equation}
\begin{split}
    G_{Ind}^{(2)}(t, 0) = \sum_{\mu\nu\gamma\epsilon}^N \gamma_{\mu\epsilon} \gamma_{\gamma\nu}  \langle \sigma_\mu^+ \sigma_\nu^+ \sigma_\gamma^- \sigma_\epsilon^- \rangle &= \gamma_0^2 \sum_{\mu\nu\gamma\epsilon}^N \delta_{\mu\epsilon} \delta_{\gamma\nu} \operatorname{Tr}\big[ \sigma_\mu^+ \sigma_\nu^+ \sigma_\gamma^- \sigma_\epsilon^- \rho(t) \big] \\
    &= \gamma_0^2 \sum_{\mu\nu}^N \operatorname{Tr}\big[ \sigma_\mu^+ \sigma_\nu^+ \sigma_\nu^- \sigma_\mu^- \rho(t) \big].
\end{split}
\end{equation}
This expression can be split into two terms, starting with $\mu=\nu$, it is evident that under the matrix multiplication, this term is identically 0, $\sigma^+ \sigma^+ = 0$. This leaves us with the terms $\mu\not=\nu$, and here we can use the fact that the total density matrix is fully factorizable to factorise the expectation value, $\langle A B \rangle = \langle A\rangle\langle B\rangle$, given that $A$ and $B$ act on different subspaces.
\begin{equation}
    \begin{split}
        G_{Ind}^{(2)}(t, 0) &= \gamma_0^2 \sum_{\mu\not=\nu}^N \operatorname{Tr}\big[ \sigma_\mu^+ \sigma_\nu^+ \sigma_\nu^- \sigma_\mu^- \rho(t) \big] \\
        &=\gamma_0^2 \sum_{\mu\not=\nu}^N \operatorname{Tr}\big[ \sigma^+ \sigma^- \rho_1(t) \big] \operatorname{Tr}\big[ \sigma^+ \sigma^- \rho_1(t) \big] \\
        & = \gamma_0^2 p^2 N(N-1).    
    \end{split}
\end{equation}
Therefore, the correlation function $\mathcal{G}^{(2)}_{Ind}(t,0) = G^{(2)}(t,0)/[G^{(1)}(t)]^2 =  (N-1)/N$. This result shows that the second-order correlation function for a set of independent emitters is $\mathcal{G}_{\rm Ind}^{(2)}(t,0) = (N-1)/N$ no matter the details of the setup. 

Using this result, in conjunction with Fig. \ref{Offset Plots}, we can state that the maximum offset correction that can be applied to the $m$-wise SM, whilst retaining the fact that the $m$-wise SM produces underestimates of the true value, is $(N-1)/N = (m-1)/m = 1/m - 1/N$.
Following the same approach for the pairwise SM, we find that in the independent limit, the method converges to the infinite emitter case $\mathcal{G}_{\rm Ind}^{(2)}(t,0) = 1$. Therefore, the offset correction, which must be subtracted from the predictions, is $1/N$.

\subsection{\textit{m}-wise SM condition} \label{Method B condition Appendix}
The critical value $N=2m$ has been shown empirically (with a margin of error). Whilst this cannot be analytically derived for the general case, it is possible to derive it when considering the specific case used by Dicke in which a collection of $N$ independent, identical emitters are found at the same location in space. This derivation is not proof of this condition being valid in any regime, but does provide evidence in support of the condition. 

The standard Dicke result (which we have shown previously) applies to all of the second-order correlation functions we have discussed: $\mathcal{G}^{(2)}$, $g^{(2)}$, $g_p^{(2)}$. Using the notation stated above, the Dicke result can be stated as 
\begin{equation}
    \mathcal{A}^{(2)}(0,0) = \frac{2(N-1)}{N}.
\end{equation}
A similar result can be obtained when considering the m-wise SM. As now there are only $m$ emitters considered, the Dicke result is 
\begin{equation}
    \mathcal{A}^{(2)}(0,0) = \frac{2(m-1)}{m}.
\end{equation}
For the pairwise method, a similar result can also be derived. For that, we start by simplifying the expressions for the denominator and numerator, respectively:
\begin{equation}
\begin{split}
    G^{(1)}G^{(1)} &= \sum_{ijkl}^2 {}^{'} A^2 \langle \sigma_i^+ \sigma_j^- \rangle \langle \sigma_k^+ \sigma_l^- \rangle \\
    &= \sum_{ijkl}^2 {}^{'} A^2 \delta_{ij} \delta_{kl} = A^2 \big( \delta_{11} \delta_{22} + \delta_{22} \delta_{11} \big) = 2A^2.
\end{split}
\end{equation}
and
\begin{equation}
\begin{split}
    G^{(2)} &= \sum_{ijkl}^2 A^2 \langle \sigma_i^+ \sigma_j^+ \sigma_k^- \sigma_l^- \rangle \\
    & = A^2 \Big( \langle \sigma_1^+ \sigma_2^+ \sigma_2^- \sigma_1^- \rangle + \langle \sigma_2^+ \sigma_1^+ \sigma_1^- \sigma_2^- \rangle + \langle \sigma_1^+ \sigma_2^+ \sigma_1^- \sigma_2^- \rangle + \langle \sigma_2^+ \sigma_1^+ \sigma_2^- \sigma_1^- \rangle \Big) \\
    & = A^2 (1-\delta_{ij})(\delta_{ik}\delta_{jl} + \delta_{il}\delta_{jk}) \\
    & = 4 A^2.
\end{split}
\end{equation}
Note that in this derivation, due to all emitters located at the same location, the coefficients $A_{ij} = A$, $\forall \ i,j$. 

So, the prediction of the pairwise SM simplifies to  $\mathcal{A}^{(2)}(0,0) = 2$. These results alone do match with the arguments stated previously, with the pairwise SM generally being an overestimate (as in this case, it is the upper bound of the Dicke result, achieved at large $N$), whereas the $m$-wise SM is a smaller value due to over-counting, and as $m \xrightarrow{} N$, this methods accuracy increases. But using these results, the error condition can be derived: we aim to find a condition in which the $m$-wise SM outperforms the pairwise SM, i.e., the error of this method is lower. This can be expressed as the inequality:
\begin{equation}
\begin{split}
    2 - \frac{2(N-1)}{N} &\geq \frac{2(N-1)}{N} - \frac{2(m-1)}{m} \\
    Nm - Nm + m &\geq Nm - m - Nm + N \\
    2m &\geq N.
\end{split}
\end{equation}

\end{document}